\documentclass[a4paper,12pt]{article}
\usepackage{amsmath}    % need for subequations
\usepackage{graphicx}   % need for figures
\usepackage{verbatim}   % useful for program listings
\usepackage{color}      % use if color is used in text
\usepackage{subfigure}
\usepackage{float}
\usepackage{amsfonts}
\usepackage{algorithmic}
\usepackage{algorithm}
\usepackage{bm}
\usepackage{multirow}
\usepackage{amsthm}
\usepackage[latin1]{inputenc}
\usepackage{tikz}
\usetikzlibrary{shapes,arrows}
\usepackage[round]{natbib}
\usetikzlibrary{shapes,arrows}
\usepackage{lscape}
\usepackage{rotating}
\usepackage{wrapfig}
\usepackage[margin=2.54cm]{geometry}
\usepackage{booktabs}
\usepackage{lineno}
\usepackage{bbm}

\usepackage[addedmarkup=nocolor]{changes}
\definechangesauthor[color=teal]{pgb}
\definechangesauthor[color=red]{mas}
\definechangesauthor[color=gray]{jlb}
\setaddedmarkup{\color{green}#1}
\newcommand{\dist}[1]{\text{#1}}

\tikzstyle{decision} = [diamond, draw, text width=6em, text badly centered, node distance=3cm, inner sep=0pt]
\tikzstyle{block} = [rectangle, draw,  text centered, minimum height=3em, text width=3cm]
\tikzstyle{be} = [rectangle, draw, text centered,rounded corners]

\tikzstyle{line} = [draw, -latex']
\tikzstyle{cloud} = [draw, ellipse,fill=red!20, node distance=3cm,
   minimum height=2em]

\setcitestyle{aysep={}}

\usepackage{authblk}

\title{\textbf{Multi-model ensembles for ecosystem prediction}}
\author[1,2,3*]{Michael A. Spence}
\author[4]        {Julia L. Blanchard}
\author[5]        {Axel G. Rossberg}
\author[6]        {Michael R. Heath}
\author[7]        {Johanna J Heymans}
\author[3,8]        {Steven Mackinson}
\author[7]        {Natalia Serpetti}
\author[6]        {Douglas Speirs}
\author[3]        {Robert B. Thorpe}
\author[1]        {Paul G. Blackwell}
\affil[1]{School of Mathematics and Statistics, University of Sheffield, Sheffield, UK}
\affil[2]{Department of Animal and Plant Sciences, University of Sheffield, Sheffield, UK}
\affil[3]{Centre for Environment, Fisheries and Aquaculture Science, Lowestoft, Suffolk NR33 0HT, UK}
\affil[4]{Institute for Marine and Antarctic Studies and Centre for Marine Socioecology, University of Tasmania, 20 Castray Esplanade, Battery Point. TAS. 7004}
\affil[5]{Aquatic Ecology Group, Department of Organismal Biology, School of Biological and Chemical Sciences, Queen Mary University of London, Mile End Road, London E1 4NS}
\affil[6]{Department of Mathematics and Statistics, University of Strathclyde, Glasgow G1 1XH, Scotland}
\affil[7]{Scottish Association for Marine Science, Scottish Marine Institute, Oban, Argyll, PA371QA}
\affil[8]{Scottish Pelagic Fishermen's Association, Heritage House, 135 - 139 Shore Street, Fraserburgh, Aberdeenshire, AB43 9BP}

\affil[*]{Corresponding author: michael.spence@cefas.co.uk}

\date{}

\begin{document}

\maketitle

\abstract{
When making predictions about ecosystems, we often have available a number of different ecosystem models that attempt to represent their dynamics in a detailed mechanistic way. Each of these can be used as simulators of large-scale experiments and make forecasts about the fate of ecosystems under different scenarios in order to support the development of appropriate management strategies. However, structural differences, systematic discrepancies and uncertainties lead to different models giving different predictions under these scenarios. This is further complicated by the fact that the models may not be run with the same species or functional groups, spatial structure or time scale. 
Rather than simply trying to select a `best' model, or taking some weighted average, it is important to exploit the strengths of each of the available models, while learning from the differences between them. To achieve this, we construct a flexible statistical model of the relationships between a collection or `ensemble' of mechanistic models and their biases, allowing for structural and parameter uncertainty and for different ways of representing reality. Using this statistical meta-model, we can combine prior beliefs, model estimates and direct observations using Bayesian methods, and make coherent predictions of future outcomes under different scenarios with robust measures of uncertainty. In this paper we present the modelling framework and discuss results obtained using a diverse ensemble of models in scenarios involving future changes in fishing levels. These examples illustrate the value of our approach in predicting outcomes for possible strategies pertaining to climate and fisheries policy aimed at improving food security and maintaining ecosystem integrity.}
\newpage

\section{Introduction}

Throughout ecology, ecosystem models are being used to support policy decisions \citep{Hyder2015,williams_2016}. Any such model is imperfect, and in order to use it to inform policy making, it is important to quantify the uncertainty of its predictions in a robust manner \citep{harwood_stokes}. In many real situations, there are several models available which each embody some knowledge of a given ecosystem, however, they often differ in their predictions. Our aim here is to describe and demonstrate a framework for using information from multiple models in a coherent way that, following \citet{chandler}, exploits their strengths and discounts their weaknesses. Our approach involves statistical modelling of the relationship between an `ensemble' of ecosystem models. To avoid ambiguity we will refer to the latter henceforth as `simulators'. 
We refer to the way in which a simulator output differs from reality as its discrepancy.

Our statistical modelling will apply Bayesian inference methods \citep{bayes_choice}, and our analysis will take into account any relevant prior knowledge as well as simulator outputs that predict what would happen in the future under different management scenarios. The Bayesian approach is subjective; for an introduction to subjective uncertainty and decision theory, see \citet{Berger}. Strictly speaking, any fully Bayesian analysis involves obtaining the posterior beliefs of a particular individual, by combining their prior beliefs with information from data and modelling. Depending on the context, that individual may be, for example, either a scientist or a policy maker. Our framework includes the elicitation of prior beliefs to combine with information from the model ensemble, allowing different individuals' posterior distributions to be obtained. For the purpose of our examples, the individual chosen in each case is one of the authors.

We first review other approaches to ensemble modelling that can be taken. One is to use a `democracy' of simulators \citep{Payne:2015aa,knutti}, 
where each simulator gets one vote, regardless of how well it represents the true system, and a distribution of possible outputs comes from this. Similarly one could take an average of the simulator outputs, which often outperforms all of the simulators \citep{Rougier}. 

However, some simulators are better at predicting some outputs better than others. 
An alternative approach is to try and find the ``best" simulator(s) \citep{Payne:2015aa,Johnson:aa}.
These methods imply that at least one of the simulators is ``correct", in the sense that it is able to predict the true output. Not only is this a bold assumption, the addition of another simulator may allow an area of the output space to become probable when before it was not. Thus by increasing the number of models there is no guarantee that the uncertainty will reduce.

One way of deciding which simulator is the ``best" is to weight simulators using Bayes factors, also known as
Bayesian model averaging \citep{Banner_Higgs,Ianelli:aa}. However, this approach depends on the likelihood of the observations given a particular simulator, so if the simulators have been fitted to different data, which is often the case in ecosystem simulators, computing Bayes factors is impossible and more ad-hoc methods are then required \citep{Ianelli:aa}. This could be further complicated as ecosystem simulators often work on different scales, giving outputs that are not directly comparable to one another.

As \citet{chandler} explains, there is generally no simulator better in all respects than the others and so there is no natural way of assigning a single weight to each simulator. 
Furthermore if simulator outputs are not presented with uncertainty then, in the case where the truth is a continuous quantity, a simulator will almost never be ``correct", thus the probability of getting the true value from the ensemble model is zero.

Climate scientists have moved away from simulator democracies and towards a more general way of weighting the simulators in an effort to keep the good parts of simulators and eliminate the bad \citep{knutti}. This leads to thinking of the outputs from a simulator as being independently sampled from a population centred on the true value \citep{tebaldi_etal2005}. In practice, there is no guarantee that the population of simulators will centre on reality and as a result simulators share biases and structural uncertainties \citep{knutti}. Furthermore, biases and discrepancies will not be independent for all simulators, as researchers who build climate simulators often contribute to a number of simulators either by developing them directly or sharing ideas with their developers. The same applies in ecology, where research groups could produce a number of ecosystem simulators, e.g. StrathE2E \citep{Heath201242} and FishSUMs \citep{Speirs2010} from University of Strathclyde, or could have similar inputs such as those coming from other simulators (e.g. European Regional Seas Ecosystem Model \citep{ERSEM}). When building an ensemble model it is important to take these similarities into account rather than treating the simulators as independent \citep{rougier_goldstein}. This has led to a number of ensemble models that treated the simulator outputs as coming from a population and explicitly modelling the difference between the consensus of the simulators and the truth \citep{tebaldi_sanso,chandler}, %. This is 
known as the shared discrepancy. 

A key assumption in these statistical models is that all the simulators represent the same dynamical process and therefore the outputs should have similar statistical structure \citep{leith}. This is not necessarily going to be the case with ecosystem models, as often their outputs are on different scales or represent different dynamical processes, which are sometimes integrated out. Furthermore these dynamics are generally less well understood than in climate science. A further difficulty in applying these methods to ecosystem simulators is that the simulators themselves have different outputs. For example in marine ecosystems, the StrathE2E simulator \citep{Heath201242} models groups of species whereas the mizer simulator \citep{blanchard} models major species individually, 
the rest of the ecosystem being included by an implicit background resources term (see appendix \ref{app:uncert} for an introduction to the simulators).  It makes sense that these simulators would, in an ensemble model, inform one another. 
For example if the StrathE2E simulator implies that the mizer simulator overestimates demersal species in general, then that suggests it is overestimating cod (\emph{Gadus morhua}) in particular and so StrathE2E is telling us something about cod indirectly.
Using this idea, a hierarchical structure for modelling the simulators allows us to sample the unobserved outputs, conditional on the simulators' observed outputs.

In this paper we describe an ensemble model which is based on the principles of \citet{chandler} but which models the outputs themselves, varying in form between simulators, rather than statistical descriptors of the outputs. 
In Section \ref{sec:simple_example} we examine a simple example of ensemble modelling by looking at the recovery times of several marine indicators.
In Section \ref{sec:dyn_model} we
setup a more general framework that will let us look at more complex examples.
In Section \ref{sec:case_study} we use the model to look at 
a specific case study: what would have happened in the North Sea if we had stopped fishing in 2013? We conclude by discussing wider applications of the approach in Section \ref{sec:dis}.

\section{Introductory example} 
\label{sec:simple_example}

We think of the available simulators as being sampled from some conceptual population of possible simulators. Our \textit{a priori} beliefs about each one are the same; we are treating them as unlabelled `black boxes'. More formally, we regard the simulators as `exchangeable'; see \citet{gelman}. We consider relaxing this assumption in Section \ref{sec:dis}.
This idea is formalised by using a hierarchical model (for more information see \citet{gelman}) 
to represent the ensemble of simulators here.

In order to demonstrate this we look at a simple example to see how long it would take indicators of good environmental status (GES) to recover if fishing was reduced \citep{GES}. 
Five simulators were run to equilibrium using fishing mortality rates representative of the period 1985-1999. The fishing mortality was then reduced by 41\%, which is the median reduction of 1985-1999 rates required to attain advised fishing mortality 
values, and the simulators run for a further 100 years. The time until each indicator recovered, defined as twice the time it took the indicator value to change halfway between the two equilibrium results, was recorded. 

The selected indicators were as follows:
\begin{itemize}
\item Seabirds and mammals biomass (B\&M): the biomass of seabirds and mammals.
\item Large fish indicator (LFI): the proportion of fish biomass pertaining to fish longer than 40cm.
\item Typical length (TyL): The biomass-weighted geometric mean length of a fish.
\item Fish population biomass trends (FPBT): Biomass of fish.
\item Ratio of zooplankton to phytoplankton (Z:P): the ratio of biomasses of zooplankton to phytoplankton.
\item Zooplankton biomass (ZB): the biomass of zooplankton.
\end{itemize}
In this example, we do not have any direct observations of the true values of the recovery times; thus we are interested in learning about the simulator consensus, $\bm\mu$.
The simulator outputs, $\bm{u}_i$, are shown in Table \ref{tb:relaxation}. We model the relationship on the log scale, with $\log_{10}\bm{u}_i=M_i\bm{x}_i$. $M_i$ is a $n_i\times6$ matrix where $n_i$ is the number of indicators output by model $i$. If model $i$ outputs the $j$th indicator, one of the rows of $M_i$ will have a 1 in the $j$th column and 0s in the other columns \citep{dominici}. $\bm{x}_i$ is a vector of the ``best guess'' of the $i$th simulator including, as latent variables, the indicators that simulator $i$ does not explicitly predict. For example, for StrathE2E,
\begin{equation*}
M_i=\begin{pmatrix}
  1 & 0 & 0 & 0 & 0 &0 \\
  0 & 0 & 0 & 0& 1 &0 \\
  0 & 0 & 0 & 0 & 0 & 1
 \end{pmatrix},
\end{equation*}
representing the fact that StrathE2E predicts the 1st, 5th and 6th indicators. The $\bm{x}_i$s are modelled as coming from a multivariate normal distribution centred on the simulator consensus, $\bm{\mu}$, with covariance $C$, 
\begin{equation*}
\bm{x}_i\sim{}\dist{N}(\bm{\mu},C).
\end{equation*}

\begin{table}[h]
\begin{center}
\caption{Predicted recovery times of UK GES indicators}
\begin{tabular}{p{3cm}ccccc}
\hline
Indicators&\multicolumn{4}{c}{Recovery time (in years)}\\
\hline
&Ecopath&FishSUMS&mizer&StrathE2E&PDMM\\
\hline
B\&M&73.4&n/a&n/a&21.5&n/a\\
LFI&4.7&7.6&5.0&n/a&8.9\\
TyL&3.7&4.6&6.1&n/a&7.4\\
FPBT&n/a&n/a&0.5&n/a&4.1\\
Z:P&2.5&n/a&n/a&3.4&1.6\\
ZB&2.7&n/a&n/a&3.2&1.6\\
\hline
\end{tabular}
\label{tb:relaxation} 
\end{center}
\end{table}

\subsection{Covariance matrix}
The covariance matrix $C$ is not known, but we can learn about it from the data, through the model for $\bm{x}_i$, and we also have relevant prior information regarding the correlations. One might expect that some of the indicators are more related than others; for example recovery times of the \emph{ratio of zooplankton to phytoplankton} and \emph{zooplankton biomass} are likely to be closely related whereas the recovery times of \emph{Seabirds and mammals biomass} and the \emph{Large fish indicator may not}. 
Given the difficulty of formulating priors on covariance matrices, we separate $C$ into the diagonal matrix $\Sigma$ 
giving the standard deviations and the correlation matrix $P$, with 
\begin{equation*}
C=\Sigma{}P\Sigma.
\end{equation*}
By applying the prior
\begin{equation*}
p(P)=\mathbbm{1}_{P\succ0}\prod_{i=1}^{n-1}\prod_{j=i+1}^n\dist{Beta}(\rho_{ij}|a_{ij},b_{ij}),
\end{equation*}
we are able to elicit experts' beliefs on correlations. Here $\mathbbm{1}_{P\succ0}$ is an indicator function that takes the value 1 if $P$ is positive definite and 0 otherwise, $\rho_{ij}$ is the element of $P$ on the $i$th row and $j$th column and $\dist{Beta}(x|a,b)$ is the density of a $\dist{Beta}(a,b)$ distribution evaluated at $x$. We also put independent scalar distributions on the diagonal elements of $\Sigma$, $\sigma_i$ defined below.

\subsection{Prior elicitation}
To get a prior distribution for a particular individual, we go through a process of `elicitation', asking them to consider a series of judgements and questions involving the parameters of the statistical model. Here, we expect to learn about $\bm{\mu}$ from the data much more readily than about $C$, so it is particularly important to elicit beliefs about $C$; we can take the prior distribution for $\bm{\mu}$ to simply be uniform. This also means that the key inferences of interest, about $\bm{\mu}$, are likely to be less sensitive to these prior beliefs.

For this example, the individual whose beliefs we focus on 
is one of the authors, MAS. He did not expect the standard deviation to be much larger than 1.0 (on the log 10 scale) as the simulators were run for a maximum of 100 years. Therefore each of the diagonal elements of $\Sigma$ were specified as
\begin{equation*}
\sigma^2_i\sim{}\dist{Exponential}(3.25),
\end{equation*}

which gives a fairly uninformative prior for $\sigma_i\in(0.25,1)$.

Using the method of concordance \citep{gokhale_press,clemen_reilly}, beta distributions were fitted to MAS' prior beliefs about the elements of the correlation matrix, $P$. MAS was asked a question of the kind: 
\begin{quote}
If two simulators were randomly chosen, and you were told that in simulator 1 the recovery time of \emph{birds and mammals biomass} was lower than in simulator 2, what is your belief that simulator 1 will have a lower recovery time for the \emph{Large Fish Indicator} than simulator 2?
\end{quote}
Using the Shelf R package \citep{oakley_shelf}, MAS' answer to this question was turned into a probability distribution. 
MAS was asked this question for each combination of indicators, giving information about inter-model correlations. This means that the prior distribution belongs to MAS and, once updated with the observations from the simulators, the posterior distribution becomes MAS' updated beliefs.

As a generally fast recovering model would recover faster than a generally slow recovering one and MAS could see no reason for it to happen the other way, 
he believed that the probability that a model with a faster recovery time for one indicator would also have a faster time for the others was almost surely above 0.5. See appendix \ref{sec:elic} for more details. MAS' prior mean for $C$ was
\begin{equation*}
\begin{pmatrix}
$0.307$ & $0.146$ & $0.147$ & $0.148$ & $0.140$ & $0.149$ \\ 
$0.146$ & $0.309$ & $0.200$ & $0.201$ & $0.165$ & $0.186$ \\ 
$0.147$ & $0.200$ & $0.306$ & $0.202$ & $0.170$ & $0.195$ \\ 
$0.148$ & $0.201$ & $0.202$ & $0.308$ & $0.170$ & $0.195$ \\ 
$0.140$ & $0.165$ & $0.170$ & $0.170$ & $0.310$ & $0.191$ \\ 
$0.149$ & $0.186$ & $0.195$ & $0.195$ & $0.191$ & $0.309$ 
\end{pmatrix}.
\end{equation*}

\subsection{Methods}
The previous sections describe the relationship between the data and the parameter of interest, $\bm{\mu}$, and the `nuisance parameter' $C$, and also the prior distributions of both parameters. Bayesian inference is then straightforward in principle, as the posterior distribution of the parameters is proportional to the product of the prior and the likelihood. In practice, the calculation of the posterior is not mathematically tractable, so we take the well established approach of using a simulation-based algorithm to sample from the posterior distribution. Because of the dimensionality and correlation of the uncertain parameter space, we fitted the model using No U-turn Hamiltonian Monte Carlo \citep{gelman_uturn} in the package Stan \citep{stan}. 

\subsection{Results}

\begin{figure}[h]
\begin{center}
\includegraphics[scale=0.7]{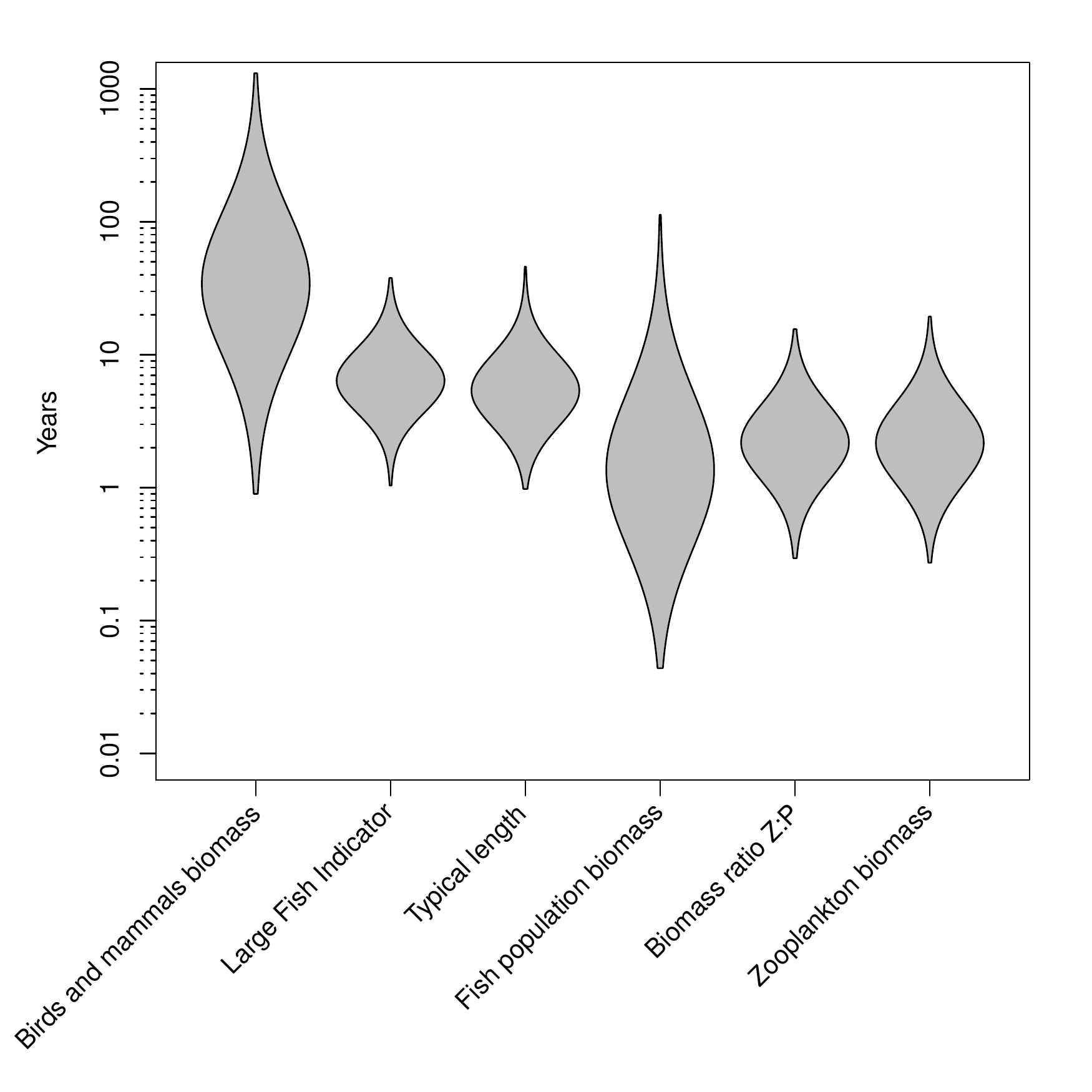}
\end{center}
\caption{The marginal posterior distributions of the simulator consensus, $\bm\mu$, for the recovery times of each of the indicators.}
\label{fig:simpleExample_mu}
\end{figure}
Figure \ref{fig:simpleExample_mu} shows the marginal posterior distributions of the elements of the simulator consensus, $\bm\mu$, of the recovery times. These represent the logically updated beliefs of MAS after learning from the observed simulator runs. MAS is more uncertain about the recovery times of the birds and mammals biomass and the fish population biomass. This is because only two of the simulators model each of these. He is much more certain about the \emph{LFI} and \emph{Typical length}, each of which was predicted by four simulators.

Despite this uncertainty, MAS has probability 0.84 that all of the indicators, except \emph{Birds and mammals biomass}, will recover within 10 years and 0.19 of recovering within five.

\section{General framework}
\label{sec:dyn_model}
In the example in the previous section we inferred the simulator consensus, $\bm{\mu}$. However, there is no reason to believe that this will be the truth \citep{chandler} so we need to allow some difference between the model consensus and truth, the shared discrepancy.

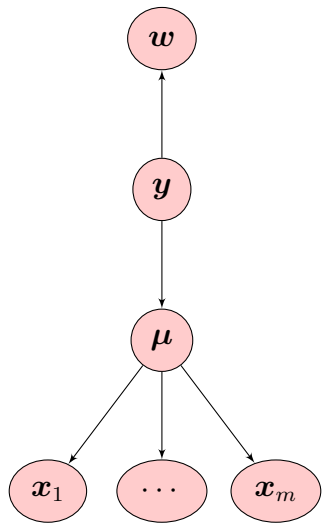
\begin{figure}[h]
\begin{center}
\begin{tikzpicture}[node distance = 4cm, auto]
    % Place nodes

\node[cloud](y_obst){$\bm{w}$};
\node [cloud,below of=y_obst,node distance=2cm] (y_t) {$\bm{y}$};
\node [cloud,below of=y_t,node distance=2cm] (mu_t) {$\bm{\mu}$};
\node [cloud,below of=mu_t,node distance=2cm] (x_2t) {$\cdots$};
\node [cloud,left of=x_2t,node distance=1.5cm] (x_1t) {$\bm{x}_1$};
\node [cloud,right of=x_2t,node distance=1.5cm] (x_3t) {$\bm{x}_m$};

  % Draw edges
  \path [line] (y_t) -- (y_obst);
  \path [line] (y_t) -- (mu_t);
  \path [line] (mu_t) -- (x_1t);
  \path [line] (mu_t) -- (x_2t);
  \path [line] (mu_t) -- (x_3t);
 
 \end{tikzpicture}
\caption{The directed acyclic graph of the toy example.}
\label{fig:sim_dag}
\end{center}
\end{figure}

To illustrate these ideas, we start by sketching out a toy example, much simpler than our actual case study. We are interested in $n$ true quantities, $\bm{y}= (y_1,\ldots,y_n)$, e.g.\ biomasses of $n$ species at a particular time. We have $m$ simulators, each giving an output representing the quantities of interest, 
$\bm{{x}}_i = (x_{i1},\ldots,x_{in})$ for $i=1,\ldots,m$.
We also have noisy observations of the truth $\bm{w}= (w_1,\ldots,w_n)$. We regard the simulators as coming from a population with mean output $\bm{\mu}=(\mu_1,\ldots,\mu_n)$, known as the simulator consensus. To define our ensemble model, we then model separately the relationships between the noisy observations and the truth, the difference between $\bm{y}$ and $\bm{\mu}$
(that is, the shared discrepancy) and the distribution of the simulator outputs around $\bm{\mu}$.
Figure \ref{fig:sim_dag} represents the ensemble model in the form of a directed acyclic graph (DAG) \citep{lunn_et_al}, where the arrows show the direct dependencies between variables.

For a realistic example, there are a number of additional factors to consider. We need to distinguish between an idealised version of simulator $i$, using the best possible parameters and inputs to produce outputs $\bm{{x}}_i$, and the available version of it with uncertain parameters producing outputs $\bm{u}_i$. This is important since we may well have information about the likely difference between $\bm{{x}}_i$ and $\bm{u}_i$, for example as a consequence of parameter estimation. Furthermore, the differences between the simulators mean that some elements of $\bm{u}_i$ may be unobserved for particular models. Finally, we are generally interested in the dynamics of the ecosystem, either for its own sake or because we are interested in future states; thus all of the above quantities are also indexed by time. Again, not all of them will actually be observed at all times; obviously we have no data corresponding directly to future times. However, in formulating the model, we retain all the corresponding variables; in particular, our aim typically is to learn about the unobserved future true values.

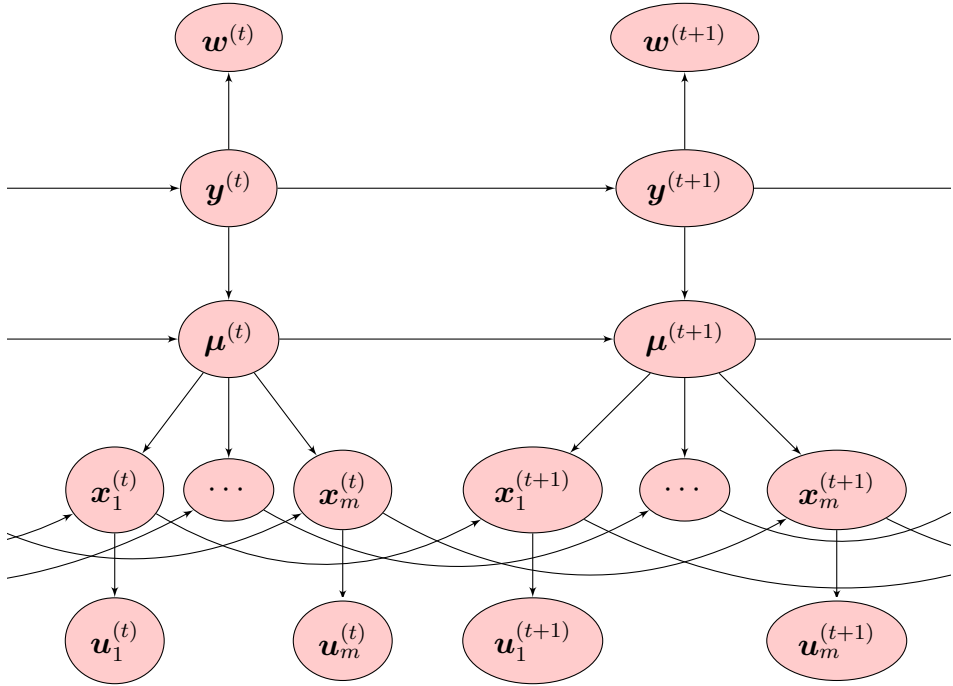
\begin{figure}[h]
\begin{center}
\begin{tikzpicture}[node distance = 4cm, auto]
    % Place nodes

\node[cloud](y_obst){$\bm{w}^{(t)}$};
\node [cloud,below of=y_obst,node distance=2cm] (y_t) {$\bm{y}^{(t)}$};
\node [cloud,below of=y_t,node distance=2cm] (mu_t) {$\bm{\mu}^{(t)}$};
\node [cloud,below of=mu_t,node distance=2cm] (x_2t) {$\cdots$};
\node [cloud,left of=x_2t,node distance=1.5cm] (x_1t) {$\bm{x}_1^{(t)}$};
\node [cloud,right of=x_2t,node distance=1.5cm] (x_3t) {$\bm{x}_m^{(t)}$};
\node [cloud,below of=x_1t,node distance=2cm] (x_1th) {$\bm{u}_1^{(t)}$};
\node [cloud,below of=x_3t,node distance=2cm] (x_3th) {$\bm{u}_m^{(t)}$};

\node [cloud,right of=y_obst,node distance=6cm] (y_obst1) {$\bm{w}^{(t+1)}$};
\node [cloud,below of=y_obst1,node distance=2cm] (y_t1) {$\bm{y}^{(t+1)}$};
\node [cloud,below of=y_t1,node distance=2cm] (mu_t1) {$\bm{\mu}^{(t+1)}$};
\node [cloud,below of=mu_t1,node distance=2cm] (x_2t1) {$\cdots$};
\node [cloud,left of=x_2t1,node distance=2cm] (x_1t1) {$\bm{x}_1^{(t+1)}$};
\node [cloud,right of=x_2t1,node distance=2cm] (x_3t1) {$\bm{x}_m^{(t+1)}$};
\node [cloud,below of=x_1t1,node distance=2cm] (x_1t1h) {$\bm{u}_1^{(t+1)}$};
\node [cloud,below of=x_3t1,node distance=2cm] (x_3t1h) {$\bm{u}_m^{(t+1)}$};

\node [below left of=x_1t,node distance=2cm] (fakey_t) {};
\node [right of=y_obst1,node distance=3.5cm] (fakey_t1) {};
 \clip (fakey_t) rectangle (fakey_t1);

\node [left of=y_t,node distance=6cm] (y_tl1){};
\node [below of=y_tl1,node distance=2cm] (fake_l1) {};
\node [below of=fake_l1,node distance=2cm] (x_1tl1) {};
\node [left of=x_1tl1,node distance=2cm] (x_2tl1) {};
\node [right of=x_1tl1,node distance=2cm] (x_3tl1) {};

\node [right of=y_t1,node distance=6cm] (y_t2){};
\node [below of=y_t2,node distance=2cm] (fake_2) {};
\node [below of=fake_2,node distance=2cm] (x_1t2) {};
\node [left of=x_1t2,node distance=2cm] (x_2t2) {};
\node [right of=x_1t2,node distance=2cm] (x_3t2) {};

  % Draw edges
  \path [line] (y_t) -- (y_obst);
  \path [line] (y_t) -- (mu_t);
  \path [line] (mu_t) -- (x_1t);
  \path [line] (mu_t) -- (x_2t);
  \path [line] (mu_t) -- (x_3t);
  \path [line] (mu_t) -- (mu_t1);
  \path [line] (mu_t1) -- (fake_2);
  \path [line] (fake_l1) -- (mu_t);
  \path [line] (y_t1) -- (y_obst1);
  \path [line] (y_t1) -- (mu_t1);
  \path [line] (mu_t1) -- (x_1t1);
  \path [line] (mu_t1) -- (x_2t1);
  \path [line] (mu_t1) -- (x_3t1);
  \path [line] (y_t) -- (y_t1);
  \path [line] (y_tl1) -- (y_t);
  \path [line] (y_t1) to (y_t2);
  \path [line] (x_1t) [bend right] to (x_1t1);
  \path [line] (x_2t) [bend right] to (x_2t1);
  \path [line] (x_3t) [bend right] to (x_3t1);
  \path [line] (x_1tl1) [bend right] to (x_1t);
  \path [line] (x_2tl1) [bend right] to (x_2t);
  \path [line] (x_3tl1) [bend right] to (x_3t);
  \path [line] (x_1t1) [bend right] to (x_1t2);
  \path [line] (x_2t1) [bend right] to (x_2t2);
  \path [line] (x_3t1) [bend right] to (x_3t2);
  
  \path [line] (x_1t) to (x_1th);
  \path [line] (x_3t) to (x_3th);
  \path [line] (x_1t1) to (x_1t1h);
  \path [line] (x_3t1) to (x_3t1h);
\end{tikzpicture}
\caption{The directed acyclic graph of the ensemble model.}
\label{fig:dag}
\end{center}
\end{figure}

Extending the notation to allow for these generalisations, we let
$\bm{y}^{(t)}$ be a vector of length $n$ of the truth at time $t$ for $t=1\ldots{}T$, where $T$ is the length of the whole simulation, $\bm{u}_i^{(t)}$ be a vector of length $n_i$ that represents the actual observed simulator outputs for $i=1\ldots{}m$, the number of simulators, and $t\in{}S_i$, the set of times at which simulator $i$ gives outputs and $\bm{w}^{(t)}$ be a vector of length $n_y$ that represents noisy observations of the truth for $t\in{}S_0$, the set of times at which there are observations. 
Let $\bm{x}_i^{(t)}$ be the unknown output from the idealised version of simulator $i$, our estimate of this will be our 
``best guess" of the
output for simulator $i$ at time $t$ where $t=1\ldots{}T$,
and let $\bm\mu^{(t)}$ be the simulator consensus at time $t$. 
We are interested in the future values of $\bm{y}^{(1:T)}$ conditional on all of the information we have received,
\begin{equation}
p(\bm{y}^{(1:T)}|\bm{w}^{(S_0:T_0)},\bm{u}_1^{(S_1:T_1)},\ldots,\bm{u}_M^{(S_M:T_M)}). 
\label{st:target}
\end{equation}
The ensemble model follows the hierarchical structure shown by the directed acyclic graph (DAG) in Figure \ref{fig:dag}.
Conditional on $\bm{y}^{(t)}$, $\bm{w}^{(t)}$ and $\bm{w}^{(t+1)}$ are independent of one another. Similarly, conditional on $\bm{x}^{(t)}_i$, $\bm{u}_i^{(t)}$ and $\bm{u}^{(t+1)}_i$ are independent of one another. The truth, $\bm{y}^{(t)}$, the model consensus, $\bm{\mu}^{(t)}$, and the simulators' ``best guesses", $\bm{x}^{(t)}_i$, do depend on their values at the previous time step. The ensemble model as a whole  is a Markov process such that conditional on the present, the past and the future are independent of one another.

Direct calculation of the distribution in equation \ref{st:target} is impossible except in the very simplest cases. In general, we use simulation-based methods such as Markov chain Monte Carlo (MCMC) to sample from 
\begin{equation*}
p(\bm{y}^{(1:T)},\bm\mu^{(1:T)},\bm{x}_1^{(1:T)},\ldots,\bm{x}_M^{(1:T)}|\bm{w}^{(S_0:T_0)},\bm{u}_1^{(S_1:T_1)},\ldots,\bm{u}_M^{(S_M:T_M)})
\end{equation*}
and hence from the distribution of interest. 
Using mathematical simplifications based on the conditional independence structure in the DAG in Figure \ref{fig:dag} (see Appendix \ref{simplify}), this approach can be implemented in standard software such as BUGS \citep{lunn_et_al}, JAGS \citep{jags} or Stan \citep{stan},
using the details of each component of the model, which are given in Sections \ref{detail:start} to \ref{detail:end}.

\subsection{The truth}
\label{detail:start}
In the absence of any simulators, our prior beliefs for the truth at time $t$, $\bm{y}^{(t)}$ follows a random walk,
\begin{equation}
\bm{y}^{(t)}=\bm{y}^{(t-1)} +\bm\epsilon_{\Lambda,t},
\end{equation}
where each $\bm\epsilon_{\Lambda,t}$ is centred on $\bm{0}$ with covariance $\Lambda_y$. At time point $t_0$, the truth, $\bm{y}^{(t_0)}$,
follows a generic prior distribution  $p(\bm{y}^{(t_0)})$.

\subsection{Direct observation}
\label{sec:obs}
At times $t\in{}S_0$, there are noisy and possibly indirect observations of the truth, $\bm{w}^{(t)}$, which come from some distribution, $p(\bm{w}^{(t)}|\bm{y}^{(t)})$ that is problem specific and is caused by data uncertainty \citep{li}. The elements of $\bm{w}^{(t)}$ may not be the same as that of $\bm{y}^{(t)}$, for example if observations are incomplete or aggregated, 
we assume that the sampling distribution of observations depends on the truth through some function $f_y(\cdot)$
such that
\begin{equation*}
\bm{\hat{w}}^{(t)}=f_{y}(\bm{y}^{(t)})
\end{equation*}
and \begin{equation*}
p(\bm{w}^{(t)}|\bm{y}^{(t)})=p(\bm{w}^{(t)}|\bm{\hat{w}}^{(t)}). 
\end{equation*}
For example if $\bm{\hat{w}}^{(t)}$ is some linear transformation of $\bm{y}^{(t)}$, then
\begin{equation*}
\bm{\hat{w}}^{(t)}=M_{y}\bm{y}^{(t)}
\end{equation*}
where $M_{\bm{y}}$ is an $n_y\times{}n$ matrix, with $n_y\leq n$ in practice.

\subsection{Model of the simulators}
The difference between the simulator consensus, $\bm\mu^{(t)}$, and simulator $i$'s ``best guess", $\bm{x}_i^{(t)}$, is simulator $i$'s individual discrepancy, $\bm{z}_i^{(t)}$, where 
\begin{equation*}
\bm{z}_i^{(t)}+\bm\gamma_i=\bm{x}_i^{(t)}-\bm\mu^{(t)}.
\end{equation*}
This distinguishes the individual discrepancy between the long-term discrepancy,
\begin{equation*}
\bm\gamma_i=\bm{\epsilon}_{\gamma,i},
\end{equation*}
where $\bm\epsilon_{\gamma,i}$ is an $n$ dimensional random variable centred on $\bm{0}$ with covariance $C$, and the short term discrepancy $\bm{z}_i^{(t)}$.
It seems natural to allow $\bm{z}_i^{(t)}$ and $\bm{z}_i^{(t+1)}$ to be dependent on each other; for example, if at time $t$, $\bm{z}_i^{(t)}$ was less than $\bm{0}$, then $\bm{z}_i^{(t+1)}$ might also be expected to be less than $\bm{0}$. With this in mind, we say that $\bm{z}_i^{(t)}$ follows a stationary auto-regressive model of order 1,
\begin{equation}
\bm{z}_i^{(t)}=R_i\bm{z}_i^{(t-1)}+\bm{\epsilon}_{z,t,i},
\label{eq:back_model}
\end{equation}
where each $\bm\epsilon_{z,t}$ is an independent $n$-dimensional random variable centred on $\bm{0}$ with covariance $\Lambda_i$ and $R_i$ is an $n\times{}n$ matrix with the constraint such that $R_i$ is stable, i.e. $\lim_{n\to\infty}R_i^n=0$. $R_i$ and $\Lambda_i$ describe the dynamics of simulator $i$ with $R_i\sim{}g_R(\bm{\theta})$ and $\Lambda_i\sim{}g_{\Lambda}(\bm{\phi})$ for some distributions $g_R$ and $g_{\Lambda}$ with hyperparameters $\bm{\theta}$ and $\bm{\phi}$ respectively. At time $t_0$, $\bm{z}^{(t_0)}_i$ is sampled from it's stationary distribution with mean $\bm0$ and covariance $\Gamma_i$, such that
\begin{equation*}
\text{vec}(\Gamma_i)=(I-R_i\otimes{}R_i)^{-1}\text{vec}(\Lambda_i)
\end{equation*}
where $\text{vec}$ is the vectorization operator representing the `stacking' of the columns of a matrix, and $\otimes$ denotes the Kronecker product of matrices.

\subsection{Uncertainty in simulator outputs}
\label{detail:end}
\label{sec:uncert_sim_out}
The simulators' outputs, $\bm{u}_i^{(t)}$, are noisy, possibly indirect, observations of the simulators' ``best guess", $\bm{x}_i^{(t)}$ for $t\in{}S_i$. The distribution of $\bm{u}_i^{(t)}$ is the posterior predictive distribution for simulator $i$. 
Furthermore, $\bm{u}_i^{(t)}$ does not necessarily contain all of the elements of $\bm{x}_i^{(t)}$. Similar to the observations of the truth, simulator $i$'s ``best guess" for the elements of $\bm{u}_i^{(t)}$ is
\begin{equation*}
\bm{\hat{u}}^{(t)}_i=f_i(\bm{x}_i^{(t)}),
\end{equation*}
and therefore $\bm{u}_i^{(t)}\sim{}p(\bm{u}_i^{(t)}|\bm{\hat{u}}^{(t)}_i)$.
In practice 
\begin{equation*}
\bm{\hat{u}}^{(t)}_i=M_i\bm{x}_i^{(t)}
\end{equation*}
is a common form.
Each simulator is fitted to a finite set of data, $D_i$, in order to find $p(\bm{\hat{u}}^{(t)}_i|D_ i)$, from which $\bm{u}_i^{(t)}$ is sampled.

\subsection{Linking the simulators and the truth}

The shared discrepancy, the difference between the simulator consensus, $\bm\mu^{(t)}$, and truth, $\bm{y}^{(t)}$, is split up into the long-term shared discrepancy, $\bm{\delta}$, and the short-term discrepancy, $\bm{\eta}^{(t)}$, i.e.
\begin{equation*}
\bm\delta+\bm{\eta}^{(t)}=\bm{y}^{(t)}-\bm\mu^{(t)}.
\end{equation*}
The short-term discrepancy is modelled with a stationary auto-regressive model of order 1
\begin{equation*}
\bm{\eta}^{(t)}=
    R_{\eta}\bm{\eta}^{(t-1)}+\bm{\epsilon}_{\eta,t},
\end{equation*}
where $R_{\eta}$ is stable and $\bm{\epsilon}_{\eta,t}$ is an $n$ dimensional random variable centred on $\bm{0}$ with covariance $\Delta$. At time $t_0$, $\eta^{(t_0)}$ is sampled from its stationary distribution with mean $\bm0$ and covariance $\Gamma_\eta$, such that
\begin{equation*}
\text{vec}(\Gamma_\eta)=(I-R_\eta\otimes{}R_\eta)^{-1}\text{vec}(\Delta).
\end{equation*}
Table \ref{tb:variables} summarises the variables in the model.

\begin{table}
\begin{center}
\caption{A summary of the variables in the ensemble model. The ensemble model is run from time 0 up until time $T$.}
\label{tb:variables}
\begin{tabular}{cll}
\hline
Variable&time period & Name \\
\hline
$\bm{y}^{(t)}$&$t=0\ldots{}T$&The truth\\
$\bm{w}^{(t)}$&$t=0\ldots{}T$&Possibly incomplete observation of truth\\
$\bm{\hat{w}}^{(t)}$&$t\in{}S_0$&Noisy observation of $\bm{w}^{(t)}$\\
$\bm\delta$&NA&Long-term shared discrepancy\\
$\bm\eta^{(t)}$&$t=0\ldots{}T$&Short-term shared discrepancy\\
$\bm\mu^{(t)}$&$t=0\ldots{}T$&Simulator concensus\\
$\bm{\gamma}_i$&NA&Simulator $i$'s long-term individual discrepancy\\
$\bm{z}_i^{(t)}$&$t=0\ldots{}T$&Simulator $i$'s short-term individual discrepancy\\
$\bm{x}_i^{(t)}$&$t=0\ldots{}T$&Simulator $i$'s best guess\\
$\bm{u}_i^{(t)}$&$t=0\ldots{}T$&Simulator $i$'s incomplete observation of $\bm{x}_i^{(t)}$\\
$\bm{\hat{u}}_i^{(t)}$&$t\in{}S_i$& Simulator $i$'s output
\end{tabular}
\end{center}
\end{table}

\section{Case Study}
\label{sec:case_study}
We illustrate our model by looking at a problem where a decision maker, who is responsible and accountable for her actions, is to make judgements about what would happen to the biomass of demersal species in the North Sea if fishing were to stop completely in 2014, using outputs from 5 different ecosystem simulators and International Bottom Trawl Survey (IBTS) data \citep{DATRAS}.  In this example, one of the authors JLB, has taken the role as the decision maker. Her prior beliefs are elicited and expressed as prior distributions and then the posterior beliefs that we show belong to her.

\subsection{Groups of species}
\label{sec:groups}
The five simulators, detailed in Appendix \ref{app:uncert}, represent demersal fish in different ways, with different species resolution and coverage. While our main interest is in demersal fish collectively, we need to represent the state of the ecosystem at a resolution that enables us to link these simulator outputs together. 

We thus group the species so that species that are represented in the same way in exactly the same simulators are in the same group, and whenever one of the simulators gives an output that is aggregated over multiple species, then that output can be expressed as a sum of one or more of our groups. 
The groups do not necessarily have any direct biological interpretation; provided the groups meet the criteria above, and allow us to represent the quantities of interest---here, demeral fish, given by the sum of all groups---the precise choice will not affect the answer obtained. 
For computational efficiency, we choose the minimum number of groups that meets these criteria while covering all demersal species. For example we grouped together monkfish, long rough dab, lemon sole and witch because they all occur in exactly the same simulators, as individual species in Ecopath and LeMans and implicitly in StrathE2E, but are not contained in any larger set of species for which this is true. This minimal set consists of 5 groups, which we will model explicitly. The groups are:
\begin{enumerate}
\item \emph{Common demersal}: 
These are cod (\emph{Gadus morhua}), haddock (\emph{Melanogrammus aeglefinus}), whiting (\emph{Merlangius merlangus}), Norway pout (\emph{Trisopterus esmarkii}), plaice (\emph{Pleuronectes platessa}), common dab (\emph{Limanda limanda}) and grey gurnard (\emph{Eutrigla gurnardus}).
\item \emph{Sole} (\emph{Solea solea}).
\item \emph{Monkfish etc.}: These are monkfish (\emph{Lophius piscatorius}), {long rough dab} (\emph{Hippoglossoides platessoides}), {lemon sole} (\emph{Microstomus kitt}) and {witch} (\emph{Glyptocephalus cynoglossus}).
%: These include: long rough dab, lemon sole, witch and monkfish.
\item \emph{Poor Cod and Rays}: These are poor cod (\emph{Trisopterus minutus}), starry rays (\emph{Amblyraja radiata}) and cuckoo rays (\emph{Leucoraja naevus}).
\item \emph{Other demersal fish}: This consists of all other demersal fish.
\end{enumerate}
We consider the total biomass densities for each of these groups, in tonnes per square kilometre, modelled on the log scale (to base 10, for ease of interpretation).

\subsection{Data and elements of the statistical model}

The IBTS data were extracted as in \citet{fung}, to reveal the total catch on the survey for each of the 5 groups for the first (1986-2013) and third quarter (1991-2013). How this value relates to the true biomass density in the North Sea is not trivial, and these values are often multiplied by catchability coefficients \citep{NDwalker_2017}  which are themselves uncertain and model-based. In this example we are only interested in the biomass density relative to 2010 and therefore the total catch from the IBTS survey is enough as we assume that catchability coefficients are constant over time. Thus each element of $\bm{y}_t$ represents the log to base 10 of the total biomass density for one of our groups of species, averaged over year $t$ year, relative to 2010. Therefore in the notation of Section \ref{sec:obs}, 
\begin{equation*}
\bm{\hat{w}}^{(t)}=f_{y}(\bm{y}^{(t)})=\bm{y}^{(t)}.
\end{equation*}
The measurement error on the observations of the truth is assumed to be normally distributed on the $\log_{10}$
scale such that
\begin{equation*}
\bm{w}^{(t)}-\bm{w}^{(2010)}\sim{}\dist{N}(\bm{y}^{(t)},\Sigma_y),
\end{equation*}
for $t\neq2010$. 
In this work we take $\Sigma_y$ to be $2\log_{10}(1.15)$ on the diagonal elements and 0 on the off diagonal elements. This was chosen so that it means that the standard deviation of the true biomass would be 15\% of the actual amount caught.

\subsection{Simulators}
\label{sec:simulators}

We have outputs from 5 different simulators, all of which have been run with zero fishing pressure from 2013 onwards.
In the next few subsections we describe the models that we used in the ensemble. 
The $i$th
simulator's output is assumed to be normally distributed on the $\log_{10}$ scale,
\begin{equation*}
\bm{u}_i^{(t)}\sim{}\dist{N}(\bm{\hat{u}}^{(t)}_i,\Sigma_i),
\end{equation*}
with $\Sigma_i$ fitted based on running simulator $i$ many times \citep{leith,chandler}. However, if this was not the case $\Sigma_i$ could be estimated within the hierarchical system. The 5 simulators and their parameter uncertainty are described in Appendix \ref{app:uncert}.

\subsection{Ensemble model}

Each element of $\bm{x}_i^{(t)}$ is the ``best guess" of simulator $i$ of the elements of $\bm{y}^{(t)}$, for $t=1968,\ldots,2100$, in log (base 10) tonnes per km of wet biomass. In this example we expect each of the simulators to converge to its own steady state, given that all external drivers are constant.
This means that in equation \ref{eq:back_model} we expect $R_i$ to tend towards 1 and $\Lambda_i$ to tend towards 0. Furthermore, if a simulator reaches a stationary state before it has stopped running, then we know that it will be in that state forever. Simulator $i$'s individual discrepancy, $\bm\gamma_i+\bm{z}_i^{(t)}$, is thus modelled as 
\begin{equation*}
\bm\gamma_i\sim\dist{N}(0,C)
\end{equation*}
and
\begin{equation*}
\bm{z}_i^{(t)}\sim{}
\begin{cases}
\dist{N}(R_i\bm{z}_{i}^{(t-1)},\Lambda_i) &\text{if }<t\leq2013,\\
\dist{N}(h_{z}(R_i,k_i,t)\bm{z}_i^{t-1},h_{\Lambda}(t,k_i)\Lambda_i) &\text{if }2014\leq{}t.
\end{cases}
\end{equation*}
where 
\begin{equation*}
h_{z}(R_i,k,t)=R_i+(1-R_i)(1-h_{\Lambda}(t,k_i))
\end{equation*}
and
\begin{equation*}
h_{\Lambda}(t,k_i)=\exp\left\{-k_i\left(t-2013\right)\right\}.
\end{equation*}
This is saying that, after the end of fishing, the variance of the truth of model $i$ reduces and the amount that the last value of $\bm{z}_i^{(t)}$ relates to the next moves towards 1 by a factor of $\exp(k_i)$ each
year.
We take $k_i\in[0,6]$, as there is not much difference numerically if $k_i$ goes above 6, with
\begin{equation*}
k_i/6\sim{}\dist{Beta}(a_k,b_k).
\end{equation*}
The diagonal elements of $R_i$ fall between $-1$ and $1$ with
\begin{equation*}
\frac{R_i+1}{2}\sim{}\dist{Beta}(\bm{a}_R,\bm{b}_R)
\end{equation*}
and the off-diagonal elements are set to 0. The model-specific variance parameter, $\Lambda_i$, is decomposed into a diagonal matrix of variances, $\Pi_i$, and a correlation matrix, $P_i$, such that
\begin{equation*}
\Lambda_i=\Pi_iP_i\Pi_i.
\end{equation*}
The form of the prior distribution for the $j$th diagonal element of $\Pi_i$ was
\begin{equation*}
\pi_{ij}\sim{}\dist{Gamma}(\alpha_{\pi,j},\beta_{\pi,j}).
\end{equation*}
Distributions over correlation matrices are complicated by the mathematical requirement of positive definiteness. In practice, we specify separate priors on the elements, and then condition on positive definiteness; the unconditional prior for the $j,k$th element of $P_i$ is given by
\begin{equation*}
\frac{\rho_{ijk}+1}{2}\sim
\begin{cases}
    \dist{Beta}(a_{\rho{}jk},b_{\rho{}jk})       & \quad \text{if } j\neq{}k,\\
    1  & \quad \text{otherwise.}\\
  \end{cases}
\end{equation*}
The difference between the truth at time $t$ and the corresponding simulator consensus, $\bm{\mu}^{(t)}$, is then
\begin{equation*}
\left(\bm{y}^{(t)}\right)-\left(\bm\mu^{(t)}-\bm\mu^{(2010)}\right)=\bm\eta^{(t)}+\bm\delta
\end{equation*}
with
\begin{equation}
\bm\eta^{(t)}\sim{}\dist{N}(R_{\eta}\bm\eta^{(t-1)},\Delta_\eta).
\label{eq:zeta_dyn}
\end{equation}
When the fishing is turned off, we are particularly uncertain about 
what will happen; thus we will remove any direct relation between $\bm{y}_t$ and $\bm{y}_{t+1}$
beyond that time. 
We will say that
\begin{equation}
\bm{\mu}^{(t)}\sim{}\dist{N}(\bm{\mu}^{(t-1)},h_{\Lambda}(t,k_{\mu})\Delta_{\mu})
\label{eq:mu_dyn}
\end{equation}
where $k_{\mu}\in[0,6]$, so that the simulator consensus reaches a stationary point, as the individual simulators do.

As in the introductory example, we focus on the subjective probabilities of a particular individual, in this case JLB. Her prior beliefs were elicited using the method described in \citet{ohaganetal} and \citet{alhussain_oakley}. Details of the prior elicitation can be found in Appendix \ref{app:prior}.

\subsection{Results}

Due to the dimensionality and correlation of the uncertain parameter space, we fitted the model using No U-turn Hamiltonian Monte Carlo \citep{gelman_uturn} in the package Stan \citep{stan}.

\begin{figure}
\begin{center}
\includegraphics[scale=0.9]{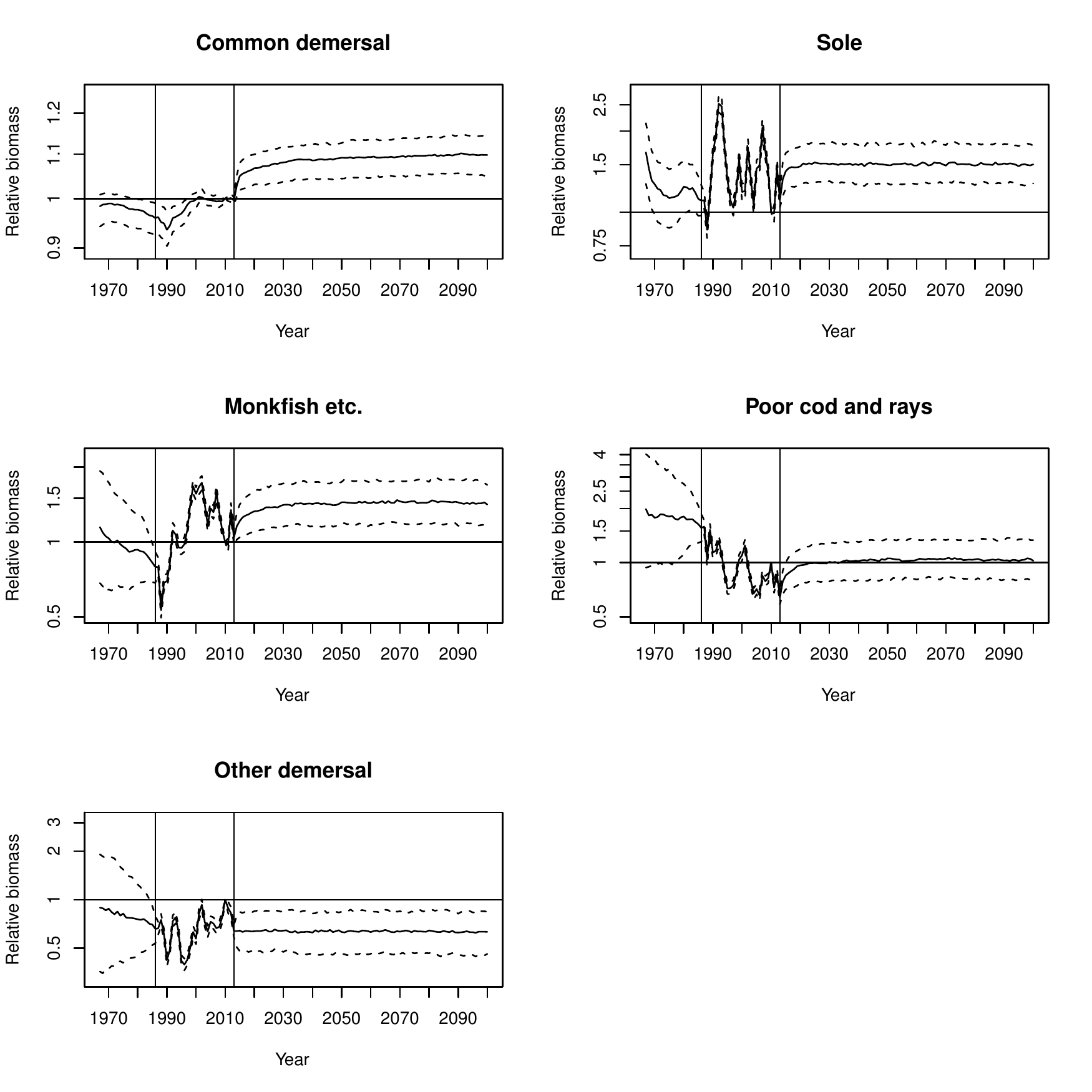}
\end{center}
\caption{Estimates of the log biomass of each group of species relative to 2010. The solid line is the median and the dotted lines are the upper and lower quartiles. The first vertical line is at 1986, the year that we first have data, and the second line is in 2013, the year before fishing were to stop completely.}
\label{fig:truth}
\end{figure}

Figure \ref{fig:truth} shows the results
for the relative
biomass over time for each group of species, if we had stopped fishing in 2013. In the notation above, that means that each plot relates to the marginal posterior distributions of each element of $\bm{y}^{(t)}$, for all $t$.
In each case, the solid line shows the posterior median output and the dotted lines the upper and lower posterior quartiles of that output. The \emph{common demersal} fish increase which is unsurprising as this group contains a lot of species targeted by fisheries and all of the individual simulators predict that.

The ensemble model and Bayesian statistical framework allow us to make probability statements, such as: the probability that there will be a greater total biomass of \emph{common demersal} in 2050 than in 2010 is 0.90. There is a similar number for \emph{sole} (0.93) and for \emph{monkfish etc. (0.88)} but it its lower for \emph{poor cod and rays} (0.55) and for the \emph{other demersal} species (0.17).

The ensemble model also `predicts' what happened before the data; that is, it gives posterior distributions for the actual values given the imperfect data and the simulator runs. Only \emph{sole} and \emph{common demersal} are output by simulators prior to 1986 and this is reflected in the increased uncertainty as we move further back in time from 1986. 

\begin{figure}
\begin{center}
\includegraphics[scale=0.9]{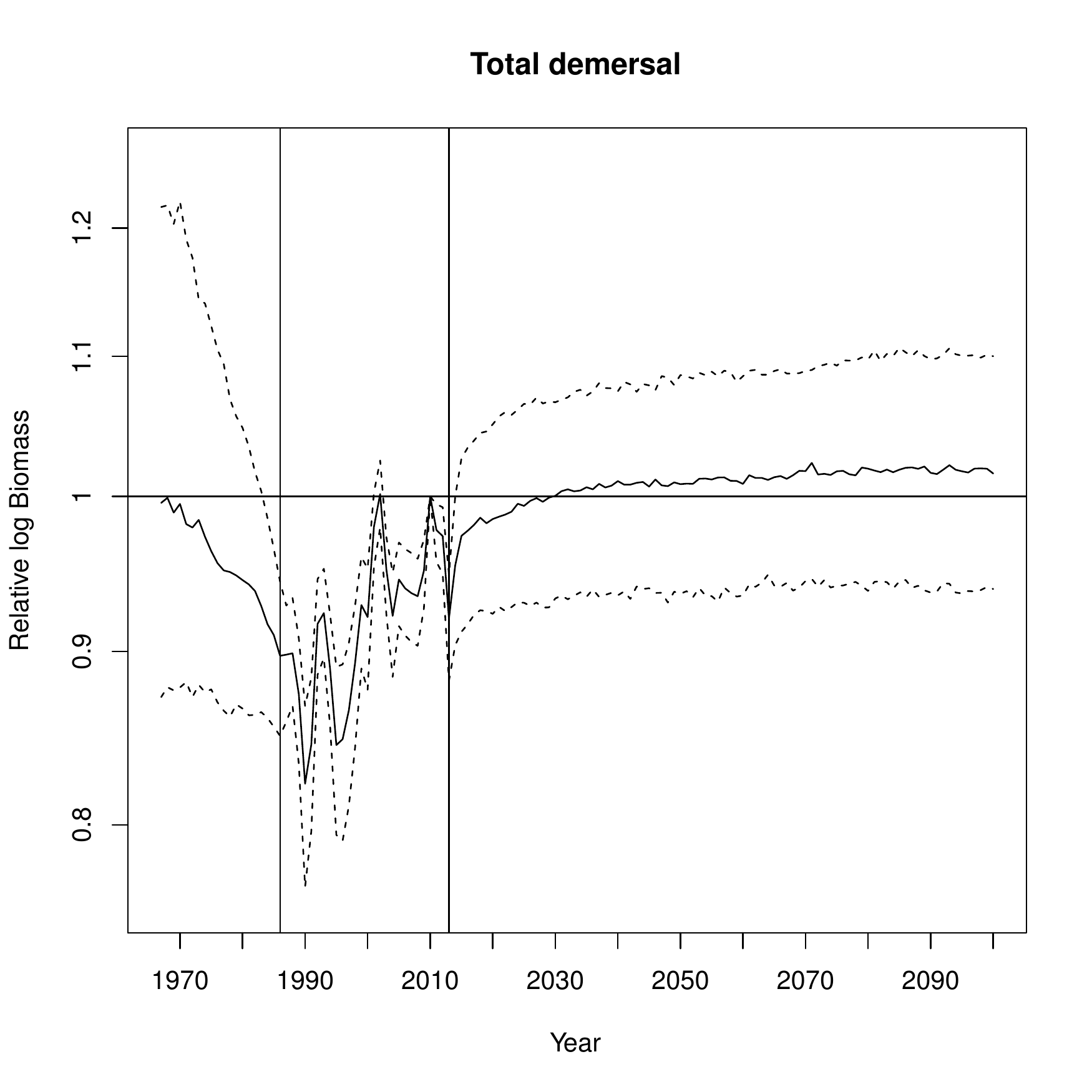}
\end{center}
\caption{The total biomass of demersal species as predicted by the models relative to 2010.}
\label{fig:tot_dem}
\end{figure}

The total biomass of demersal species is difficult to calculate here because the discrepancy between the 
simulator consensus
and the truth is difficult to quantify. 
We do not have direct survey data the we can use for true total demersal biomass; values depend on the varying, and unknown, catchability coefficients for each of the groups. Figure \ref{fig:tot_dem} shows the total demersal biomass if we assumed that the groups had the same catchability coefficient. It shows that we are rather uncertain about whether the biomass will grow relative to the biomass in 2010. However, what it was before 1986 is quite uncertain. This is because of the uncertainty in the populations of \emph{Other demersal species}. We found that in 2050 the biomass will be larger than in 2010 with probability 0.55.

\begin{figure}
\begin{center}
\includegraphics[scale=0.9]{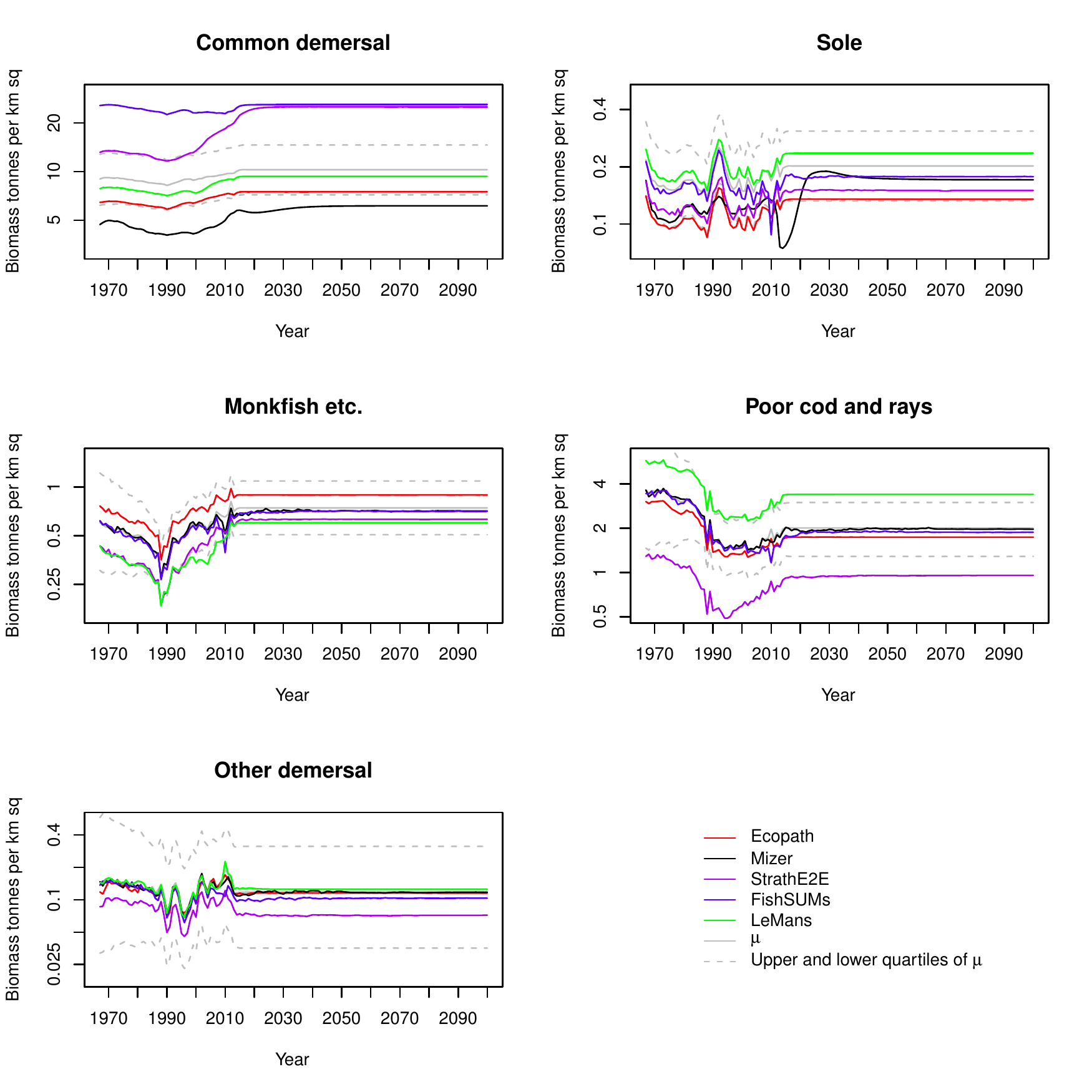}
\end{center}
\caption{The median best guess for the simulators ($\bm{x}_i$) for mizer (black), FishSUMS (purple), LeMans (green), Ecopath (red) and StrathE2E (pink) and the median simulator consensus ($\bm\mu$) and its quartiles in solid grey and dotted grey respectively.}
\label{fig:mean_and_run}
\end{figure}

The median ``best guess" of each of the simulators is shown in Figure \ref{fig:mean_and_run}. Notice that StrathE2E predicts quite a large increase in \emph{common demersal} despite not explicitly outputting it.

The posterior predictive distribution for the relative truth in 2025 for \emph{common demersal} and \emph{monkfish etc.}\ is positively correlated (0.28). This suggest that learning something about the \emph{common demersal} would tell you something about \emph{monkfish etc}. Hence the mizer simulator gives some information regarding the \emph{monkfish etc.}\ despite not actually predicting it. See Appendix \ref{app:res} for the other correlations between the groups.

\section{Discussion}
\label{sec:dis}

By treating the simulator outputs as coming from a population of simulators and modelling this population, we have presented in this paper a general way of combining ecosystem simulators in order to inform a decision maker about the forecast under a specific management strategy. Our model combines a number of different simulators, exploiting their strengths and discounting their weaknesses \citep{chandler} to best inform the decision maker.

\subsection{Case study results}

We demonstrated how to combine simulators with different outputs by predicting how fast different indicators would recover if we were to reduce the fishing mortality.
We further demonstrated our model by using 5 ecosystem simulators and investigating what would have happened to demersal fish in the North Sea if we had stopped fishing in 2014. 
We found that although the total biomass of demersal fish may not increase over time, the biomass of targeted fish will likely increase relative to 2010. 

\subsection{General model features}
One of the difficulties in building an ensemble model with ecosystem simulators is that the simulators outputs are often done on different scales and are not directly comparable, for example StrathE2E models groups of species (e.g. pelagic, demersal) whereas mizer models major species individually. 
Our approach, unlike existing methods  of combining simulators (e.g. Bayesian model averaging \citep{Banner_Higgs,Ianelli:aa}), allows us to combine outputs from these widely differing simulators. We achieve this by modelling what each simulator would predict for each of the groups of species we are interested in, whether it is explicitly modelled or not by the simulator.

For example, in the case study, StrathE2E only models the total demersal species. Using information from the other simulators regarding the breakdown of demersal species and how the dynamics between species work, the ensemble model is able to say what StrathE2E would predict on a species level. 

Ecopath and StrathE2E both predict sums of things. For Ecopath it is the sum of \emph{poor cod and rays} and \emph{other demersal} and StrathE2E gives the sums of all of the groups. As with the simulators that do not predict specific species, we are able to infer what these models predict about the things that they sum over though correlations learned by other simulators. In this sense, the mizer model, which only predicts \emph{common demersal} and \emph{sole}, gives information about how StrathE2E divides its demersal species and therefore gives some information to other species. Therefore, if we were interested in what would happen to the other demersals if we were to stop fishing, we should include all of the simulators despite only two of them predicting it.

The ensemble enables the uncertainty of the predictions to be quantified in a robust manner. The uncertainty in the prediction increases the further away from the observations of the truth both when forecasting and hindcasting. The uncertainty increases when there are fewer simulators that give outputs. All of the simulators give outputs for the \emph{common demersal}, four explicitly and one implicitly, and therefore we are more certain about what will happen in the future whereas for \emph{poor cod and rays}, where only three simulators predict values for the future and only one explicitly, the uncertainty is much higher. The uncertainty is highest for other demersal species. This is understandable as only two simulators predict values for this group of species, neither of which does so explicitly.

The hierarchical distributions for the covariance parameter for the short-term individual discrepancy, $\Lambda_i$, was divided into a diagonal matrix $\Pi_i$, with the $j$th diagonal element being the standard deviation of the $j$th element, and correlation matrix $P_i$. It was divided this way, as there is more information about the dynamics of species in models as opposed to the variance. In ecosystems simulators, the dynamics are going to be similar in direction but maybe not in magnitude.

In the case study, we used beta distributions for each of the off diagonal elements of the correlation matrix and then conditioned on positive definiteness. This enabled us to learn about each element of the correlation matrix separately which is not possible in other formulations of the covariance matrix \citep{Alvarez_et_al}. It was also important to use informative priors as none of the simulators explicitly model \emph{other demersal}. As there is no lower bound (on the log scale) for the values of the ``best guess" of \emph{other demersal}, we required some prior information about the distribution of the standard deviations, $\Pi$. This does suggest that the ensemble prediction is somewhat based on that of the priors for $\Lambda_i$. In practise we suggest checking your ensemble model predicts in a way that the decision maker believes before data observing the truth. In the case study described here that is prior to 1986, similar to the hypothetical data method of \citet{Kadane}.

When building the ensemble model, how the species groups are decided depends on the question being asked. In the case study we were interested in what would happen to demersal fish if we were to stop fishing, so we grouped the species into as few groups as possible. However, if we were interested in another question, for example if we had been interested in what would happen to commercial fish, we would divide the species into groups with commercial and non-commercial fish conditioned on species in each group being presented in exactly the same simulators. As the number of groups increases,
the dimensions of the covariance matrices increase, so we advice that the number of groups be kept to a minimum as this would aid computation time and require less simulators and prior elicitation.

Using the ensemble model developed here, there is no need to identify the ``best model" driven by the question being asked \citep{Dickey-Collas:2014aa}, but one should include all available simulators. Rather than developing a number of simulation models to answer different specific questions, the ensemble model can be designed to answer the question at hand. Furthermore, as the ensemble model aims to exploit simulators strengths and discount their weaknesses, it is better for a simulator to be really good at modelling one aspect of the ecosystem than being okay at modelling a lot of things.

Due to the nature of the different ecosystem simulators, they often have different processes and are often unable to run the same scenarios, for example the mizer model doesn't have climate dynamics included in it.
If we are interested in one of the scenarios that a specific simulator is unable to run we should still include that simulator in the ensemble model as it gives information about how species interact with one another as well as the state of the ecosystem up until the current time. In order to include this simulator in the ensemble, we could  increase $\Sigma_i$ as a function of time with in future scenarios. This would suggest mean that the simulators ``best guess", which in this case means what the simulator would predict if it were able to run the scenario, would be less informed by the simulator output as time went on.

\subsection{Future work and extensions}

Some ecosystem simulators are more similar than others, for example there are a number of size-based simulators in the marine literature \citep[e.g.][]{blanchard_2009,mizer} that are very similar, which may violate the exchangeability assumption made in Section \ref{sec:dyn_model}. Additional hierarchy could be added to the ensemble model that would allow such simulators to have more similar discrepancies. 

In climate science, where the simulators are very similar to one another and it is possible to create a phylogenetic tree \citep{knutti_2013} that shows the development history of each simulator, \citet{Demetriou_phd}
did add additional hierarchy allowing closely related simulators to have similar discrepancies. They found that the major source of uncertainty was that of the shared discrepancy and that the results of the ensemble model were very much similar to all of the simulators being exchangeable. 

Additionally, if there were multiple observations, it is possible to include this by adding a number of observations of the real system $\bm{y}$. This could be important in ecology, as it can often be difficult to get a direct observation of something of interest. Modelling $\bm{y}$ by including multiple observations could be a way to learn about things that we are unable to observe and therefore learn about the shared discrepancy.

In this paper, we have demonstrated the ideas and methods in cases where the quantities of interest are of fairly low dimension and have joint Gaussian distributions. However, with the increased efficiency of new statistical software and algorithms \citep[see e.g.][]{girolami}, it is possible to address larger problems involving more general distributions.

\subsection{Conclusion}

This work brings ecology on track to synthesise work or many modelling studies that have been and are being conducted in such a way that we can obtain more holistic knowledge over a wide scope of complex ecological systems, including a clearer, quantitative understanding uncertainties and knowledge gaps. This enables us to make coherent forecasts that take into account all that we have learnt from the simulators collectively.

\section*{Acknowledgments}
The work was supported by the Natural Environment Research Council and Department for Environment, Food and Rural Affairs [grant number NE/L003279/1, Marine Ecosystems Research Programme]. The authors would like to thank Tom Webb, Remi Vergnon, Yuri Artioli, S\'{e}vrine Saillery, Paul Somerfield, Melanie Austen, Nicola Beaumont and Stefanie Broszeit for participating in early elicitation exercises.

\section*{Author contribution}
MAS, PGB and JLB conceived the ideas and designed the methodology; 
AGR conceived and co-ordinated the introductory example;
JLB extracted the data for the main case study; 
MAS, MRH, SM, DS, AGR, RBT, JJH and NS ran the simulators for the different scenarios; 
MAS implemented the methodology; 
MAS and PGB analysed the data; 
MAS and PGB led the writing of the manuscript. 
All authors contributed critically to the drafts and gave final approval for publication.

\appendix

\renewcommand{\thefigure}{A\arabic{figure}}
\setcounter{figure}{0}
\renewcommand{\thetable}{A\arabic{table}}
\setcounter{table}{0}

\section{Conditional independence structure of the ensemble model}
\label{simplify}
In order to implement the ensemble model, we can write
\begin{eqnarray}
&&p(\bm{y}^{(1:T)},\bm\mu^{(1:T)},\bm{x}_1^{(1:T)},\ldots,\bm{x}_M^{(1:T)}|\bm{w}^{(S_0:T_0)},\bm{u}_1^{(S_1:T_1)},\ldots,\bm{u}_M^{(S_M:T_M)})\nonumber\\
&&\quad\propto{}p(\bm{y}^{(1:T)})p(\bm{w}^{(S_0:T_0)}|\bm{y}^{(1:T)})p(\bm\mu^{(1:T)}|\bm{y}^{(1:T)})\nonumber\\
&&\qquad\times{}p(\bm{x}_1^{(1:T)},\ldots,\bm{x}_M^{(1:T)}|\bm\mu^{(1:T)},\bm{y}^{(1:T)})\nonumber\\
&&\qquad\times{}p(\bm{u}_1^{(S_1:T_1)}\ldots\bm{u}_M^{(S_M:T_M)}|\bm\mu^{(1:T)},\bm{y}^{(1:T)},\bm{x}_1^{(1:T)},\ldots,\bm{x}_M^{(1:T)}).
\label{eq:full}
\end{eqnarray}
Using the conditional independence structure in the DAG we can simplify equation \ref{eq:full} to
\begin{eqnarray}
p(\bm{y}^{(t_0)})
p(\bm{w}^{(t_0)}|\bm{y}^{(t_0)})^{\mathbb{I}_{t_0\in{}S_0}}p(\bm\mu^{(t_0)}|\bm{y}^{(t_0)})
\prod_{i=1}^m\Bigg\{
p(\bm{x}_i^{(t_0)}|\bm\mu^{(t_0)})p(\bm{u}_i^{(t_0)}|\bm{x}^{(t_0)}_i)^{\mathbb{I}_{t_0\in{}S_i}}
\nonumber\\
\prod_{t=t_0-1}^1
p(\bm\mu^{(t)}|\bm{y}^{(t)},\bm\mu^{(t+1)})p(\bm{y}^{(t)}|\bm{y}^{(t+1)})p(\bm{w}^{(t)}|\bm{y}^{(t)})^{\mathbb{I}_{t\in{}S_0}}
p(\bm{x}_i^{(t)}|\bm\mu^{(t)},\bm{x}_i^{(t+1)})p(\bm{u}_i^{(t)}|\bm{x}_i^{(t)})^{\mathbb{I}_{t\in{}S_i}}
\nonumber\\
\prod_{t=t_0+1}^T
p(\bm\mu^{(t)}|\bm{y}^{(t)},\bm\mu^{(t-1)})p(\bm{y}^{(t)}|\bm{y}^{(t-1)})p(\bm{w}^{(t)}|\bm{y}^{(t)})^{\mathbb{I}_{t\in{}S_0}}
p(\bm{x}_i^{(t)}|\bm\mu^{(t)},\bm{x}_i^{(t-1)})p(\bm{u}_i^{(t)}|\bm{x}_i^{(t)})^{\mathbb{I}_{t\in{}S_i}}\Bigg\}.
\label{eq:con_ind}
\end{eqnarray}
In practice, standard MCMC software enables us to sample from the model simply by specifying each of the components in equation \ref{eq:con_ind}, as in Sections \ref{detail:start} to \ref{detail:end}.

\section{Complete MAS elicitation}
\label{sec:elic}
As a generally fast recovering model would on average recover faster than a generally slow recovering one and MAS could see no reason for it to happen the other way, he said that he believe that the probability that a model with a faster recovery time for one indicator would also have a faster time for the others was almost surely above 0.5.

Seabirds and mammals are long lived and therefore there dynamics will be slower than shorter lived species. Thus their recovery times will be much larger than fish and plankton. The recovery times will very much depend on the model in question and not massively linked to that of the other indicators. However, as mentioned above, a model that recovers quickly for a few indicators is likely to recover quickly for them all and therefore MAS said that he would expect a median value of 0.75, for the proportion of models that predicted a faster recovery of birds and mammals biomass to also predict faster recoveries of the other indicators with quartiles of 0.65 and 0.8.

He believed that there is a much stronger link between the LFI and the typical length. As these two are measuring similar things with large fish having a disproportionate effect on the typical length than smaller fish, a recovery in one would imply a recovery in another. Therefore MAS predicted that a faster recovery in LFI in one model would mean that there was a strong probability of a faster recovery of the typical length in the same model. There is also a similar relationship for fish population with LFI and typical length.

MAS was more uncertain about the relationship between LFI and the ratio of zooplankton and phytoplankton. As with all of the other relationships, he believed that the proportion of models would be larger than 0.5 but was unsure to what extent. This relationship held for typical length and fish population biomass with ratio of zooplankton and phytoplankton.

There is a link between the LFI, typical length and fish population biomass with the biomass of zooplankton. These indcations are dynamically correlated and therefore the recovery of one means that the second will recover quickly. Thus, the proportion will be high but not too high as the time it took for the zooplankton to filter through to the fish is model dependent.

The ratio of zooplankton and phytoplankton and zooplankton biomass will recover in similar times as for one to recover, the zooplankton biomass needs to recover in both. Therefore these will have a strong relationship.

\section{Simulators}
\label{app:uncert}

\subsection{Multispecies size spectrum model}
\label{sec:mort_mizer}
The multispecies size spectrum model (mizer) was developed to represent the size and abundance of all organisms from zooplankton to large fish predators in a size-structured food web. A proportion of the organisms are represented by species specific-traits and body size while others are represented solely by body size. In this form, the model has principally been used to describe the effects of fishing on interacting species and the size-spectrum. 

Mizer provides predictions of the abundance of each species at size. The core of the model involves ontogenetic feeding and growth, mortality, and reproduction driven by size-dependent predation and maturation processes \citep{hartvig,mizer}. It thus differs from some other size-based models that assume deterministic growth based on life history parameters. The smallest individuals in the model do not eat fish belonging to the fish populations, but consume smaller planktonic or benthic organisms which we describe as a background resource spectrum. Fish grow and die according to size-dependent predation and, if mature, recruit new young which are put back into the system at the minimum weight. The model is able to predict abundance at size, biomass, growth and mortality rates for each species. For a complete description of the model see \citet{hartvig} or \citet{mizer}.

\citet{blanchard} developed and applied a version of mizer for the North Sea. In the model, 12 of the more common species have been explicitly represented. It is this version of mizer that has been used in this study.

\subsubsection{Introductory example}
Mizer was able to predict recovery times for the Large fish indicator (LFI), Typical length (TyL) and Fish population biomass trends (FPBT). Therefore
\begin{equation*}
M_{mizer}=\begin{pmatrix}
  0 & 1 & 0 & 0 & 0 & 0\\
  0 & 0 & 1 & 0 & 0 & 0 \\
  0 & 0 & 0 & 1 & 0 & 0
 \end{pmatrix}.
\end{equation*}

\subsubsection{Case study}
The simulator is able to run from 1968 until 2100 and explicitly outputs  \emph{common demersal} and \emph{sole}, therefore
\begin{equation*}
\bm{\hat{u}}^{(t)}_{mizer}=f_{mizer}(\bm{x}_{mizer}^{(t)})=M_{mizer}\bm{x}_{mizer}^{(t)},
\end{equation*}
with
\begin{equation*}
M_{mizer}=\begin{pmatrix}
  1 & 0 & 0 & 0 & 0 \\
  0 & 1 & 0 & 0& 0
 \end{pmatrix}
\end{equation*}
for $t=1968\ldots2100$. 

We use parameter values from \citet{spence_et_al_2015} to simulate up until 2010 and, assuming conditionally independent Gaussian errors on the landings, we used a particle filter (see \citet{smc15year} for an introduction to particle filters) to update the fishing mortalities for 2011-13. 100 samples from the joint posterior distribution were simulated from 1968-2100 with fishing being turned off in 2013. 

\subsection{Ecopath}

Ecopath was developed first in 1984 by \citet{polovina} and has been updated subsequently to include temporal (Ecosim) and spatial (Ecospace) dynamics \citep{christensen_walters} and is currently used extensively to simulate historic changes in ecosystems \citep{HEYMANS2016173}. The Ecopath model used in this case is the model of the North Sea \citep{lynam_mackinson}. It contains $>10$ fishing fleets and $>60$ functional groups and  some of which are split into multiple age stanzas.

\subsubsection{Introductory example}
Ecopath was able to predict recovery times for Birds and mammals (B\&M), the Large fish indicator (LFI), Typical length (TyL), Zooplankton to phytoplankton biomass (Z:P) and Zooplankton biomass (ZB). Therefore
\begin{equation*}
M_{EwE}=\begin{pmatrix}
  1 & 0 & 0 & 0 & 0 & 0\\
  0 & 1 & 0 & 0 & 0 & 0\\
  0 & 0 & 1 & 0 & 0 & 0 \\
  0 & 0 & 0 & 0 & 1 & 0\\
  0 & 0 & 0 & 0 & 0 & 1
 \end{pmatrix}.
\end{equation*}

\subsubsection{Case study}
In this example, the simulator is a able to run from 1991 to 2023 and explicitly predicts: \emph{common demersal}, \emph{sole}, \emph{monkfish etc.}\ and the sum of \emph{poor cod and rays} and \emph{other demersal}. Although the $\bm{x}$s are on the $\log_{10}$ scale, we have to transform them onto the real scale in order to add them. Therefore
\begin{equation*}
\bm{\hat{u}}^{(t)}_{EwE}=f_{EwE}(\bm{x}_{EwE}^{(t)})=\log_{10}\left(M_{EwE}10^{\bm{x}_{EwE}^{(t)}}\right),
\end{equation*}
with
\begin{equation*}
M_{EwE}=\begin{pmatrix}
  1 & 0 & 0 & 0 & 0 \\
  0 & 1 & 0 & 0& 0 \\
  0 & 0 & 1 & 0 & 0 \\
  0 & 0 & 0 & 1 & 1 
 \end{pmatrix}
\end{equation*}
for $t=1991-2023$.

\subsection{FishSUMS}
\label{sec:mort_fishsum}
The FishSUMS model \citep{Speirs2010,Speirs2016} represents the population dynamics of a set of key trophically-linked predator and prey species. For each species the state variables are biomass by length class. In discrete time steps the state variables are updated through increasing length, density-dependent mortality, and losses due fishing and predation by explicitly modelled species, and seasonal reproduction. Additional food resources, not modelled at the species level, are characterised by three biomass spectra representing zooplankton, benthos, and ``other fish''. Outputs from the model are time series of total species biomass (TSB), normalised length distributions at annual census dates, annual recruitment, catch and landings, for each of the focal species. 

The model was initially configured for the North Sea with a set of nine structured species focused on cod and its main predators and prey \citep{Speirs2010}, and subsequently extended to include plaice and saithe so as to include the eight most abundant demesral species that make up $>90\%$ of the North Sea biomass \citep{Speirs2016}. In general the model is configurable for any set of structured species and unstructured prey groups. The model has been developed as a package for the R software environment, available on request.

\subsubsection{Introductory example}
FishSums was able to predict recovery times for the Large fish indicator (LFI) and Typical length (TyL). Therefore
\begin{equation*}
M_{FishSUMS}=\begin{pmatrix}
  0 & 1 & 0 & 0 & 0 & 0\\
  0 & 0 & 1 & 0 & 0 & 0
 \end{pmatrix}.
\end{equation*}

\subsubsection{Case study}

The simulator is able to run from 1990 to 2098 and explicitly models the \emph{common demersal} only, so
\begin{equation*}
\bm{\hat{u}}^{(t)}_{FishSUMS}=f_{FishSUMS}(\bm{x}_{FishSUMS}^{(t)})=M_{FishSUMS}\bm{x}_{FishSUMS}^{(t)},
\end{equation*}
with
\begin{equation*}
M_3=\begin{pmatrix}
  1 & 0 & 0 & 0 & 0 
 \end{pmatrix}
\end{equation*}
for $t=1990-2098$.

The simulator was ``hand tuned" \citep{Speirs2010} to fit observed data in order to find the optimal parameter set for the model. During this period it became apparent that the most sensitive parameters were the mortality rate parameters. For this work we fitted these 4 parameters to landings data from 1990 to 2008. We assumed that the true landings were normally distributed on the log scale \citep{tsechay} and then fitted these parameters using an MCMC sampler \citep{metropolis,hastings} to landings between 1990 and 2008.

For the best point in the MCMC algorithm we fitted the fishing mortalities from 2008 until 2013 using a particle filter. 
We ran 100 parameter sets found using the Metropolis-within-Gibbs sampler with fishing mortalities fitted to 2008-2013, and ran it from 1991-2090 with fishing being turned off in 2013.

\subsection{Strathclyde End to End}

The Strathclyde end-to-end  (StrathE2E) marine food web model was designed to simulate regional scale, macroscopic top-down and bottom-up cascading trophic effects \citep{heath_eco_let}. The mathematical formulation is based on a network of coupled ordinary differential equations representing the entire food web in the water column and seabed sediments from nutrients and microbes though zooplankton and fish, to birds and mammals, including the effects of advection, mixing and active vertical migrations. Living components are represented at low taxonomic resolution, focussing on fluxes of nitrogen between coarse functional groups, and simulating the general ?shape? of the food web rather than the detail. The scheme takes off-line output from General Circulation Models (GCM) in the form of volume and nutrient fluxes through the external boundaries and mixing rates between the vertical compartments, but is not directly coupled to any GCM. The advantage is very fast run-times which has enabled the implementation of computational parameter optimisation methods to fit the models to observed data, sensitivity analysis \citep{morris_spires}, and computation of likelihoods for model outputs. The focus of existing uses of Strathe2E has been on UK shelf seas and the cascading implications of fisheries and fishing practices such as trawling and its impacts on the seabed, and discarding of unwanted catch \citep{Heath201242,heath_nature,heath_scot_rep}.

\subsubsection{Introductory example}
StrathE2E was able to predict recovery times for b=Birds and mammals (B\&M), Zooplankton to phytoplankton biomass (Z:P) and Zooplankton biomass (ZB). Therefore
\begin{equation*}
M_{StrathE2E}=\begin{pmatrix}
  1 & 0 & 0 & 0 & 0 & 0\\
  0 & 0 & 0 & 0 & 1 & 0\\
  0 & 0 & 0 & 0 & 0 & 1
 \end{pmatrix}.
\end{equation*}

\subsubsection{Case study}
The simulator is able to run from 1983 until 2050 and is able to output a sum of the different groups of species. Although the $\bm{x}$s are on the $\log_{10}$ scale, we have to transform them onto the real scale in order to add them so
\begin{equation*}
\bm{\hat{u}}^{(t)}_{StrathE2E}=f_{StrathE2E}(\bm{x}_{StrathE2E}^{(t)})=\log_{10}\left(M_{StrathE2E}10^{\bm{x}_{StrathE2E}^{(t)}}\right),
\end{equation*}
with
\begin{equation*}
M_{StrathE2E}=\begin{pmatrix}
  1 & 1 & 1 & 1 & 1 \\
 \end{pmatrix}
\end{equation*}
for $t=1983-2050$.

In \citet{Heath201242}, the simulator was fitted to data using a method of simulated annealing.
For this work we took the parameter values from the simulated annealing and then used priors from \citet{morris_spires} to find the parameters that are most sensitive. 
Using derivative-based sensitivity analysis \citep{dbgsm,dbgsa1} we calculated upper bounds of the total sensitivity indices for each of the 72 uncertain parameters.
We then used MCMC in order to sample from the posterior distribution of the 12 most sensitive parameters.

We sampled 100 parameter values from the posterior distribution and then, using the fishing mortalities from the multi species model and the FishSUMs, calculated in Sections \ref{sec:mort_mizer} and \ref{sec:mort_fishsum}  and calibrated to the StrathE2E. We ran StrathE2E to 2013 and then the fishing mortalities were set to 0 before running it until 2050.

\subsection{LeMans}
The LeMans North Sea model framework \citep{thorpe,thorpe_16,thorpe_17} is an ensemble of length-structured multispecies models which account for multispecies interactions and model parameter uncertainty. It is a modified form of the length-based multispecies model initially developed by \citet{hall} to represent the Georges Bank fish community, and which was subsequently adapted for use in the North Sea by \citet{rochet_etal}. The model represents 21 fish species in 32 equal length classes of around 5cm each, spanning the full size range of species represented into the model (nearly 200cm for some simulations). Progression of individuals through length classes is represented by a deterministic von Bertalanffy growth equation. Individuals mature when they reach a certain size which is defined by a logistic model, in which 50\% of the individuals maturing at the length of maturity ($Lmat$  see Table S2 in \citet{thorpe_17}). Reproduction is described with a spawner recruit relationship, which determines the numbers of recruits entering the smallest size class from the biomass of mature individuals. Species dynamics are linked via predation mortality (M2) which varies with predator abundance, and size and species preference. Size preference is described with a preference function based upon a log-normal distribution and species preference with a diet matrix indicating who eats whom \citep{rochet_etal,thorpe}. In each length class, individuals are also susceptible to residual natural mortality (M1) and fishing mortality (F). An ensemble approach is used, based upon a ``filtered ensemble" (FE) of models drawn from a population of 78,125 candidate models (the ``unfiltered ensemble'' or UE), with the FE being selected on the basis of an individual member's ability to persist stocks when unfished, and to simulate assessed abundances of 10 stocks between 1990 and 2010 to an acceptable degree. This ensemble approach is described in detail in \citet{thorpe}, and further details of the model, including equations are provided in \citet{thorpe_17}.

\subsubsection{Case study}
The simulator was able to predict from 2000 until 2099 and explicitly models: \emph{common demersal}, \emph{sole}, \emph{monkfish etc.}\ and \emph{poor cod and rays}, therefore
\begin{equation*}
\bm{\hat{u}}^{(t)}_{LeMans}=f_{LeMans}(\bm{x}_{LeMans}^{(t)})=M_{LeMans}\bm{x}_{LeMans}^{(t)},
\end{equation*}
with
\begin{equation*}
M_{LeMans}=\begin{pmatrix}
  1 & 0 & 0 & 0 & 0 \\
  0 & 1 & 0 & 0& 0 \\
  0 & 0 & 1 & 0 & 0 \\
  0 & 0 & 0 & 1 & 0 
 \end{pmatrix}
\end{equation*}
for $t=2000\ldots2099$.

\subsection{PDMM}

The Population-Dynamical Matching Model (PDMM) constructs complex and population-dynamically stable ecological model communities at species resolution by mimicking the natural process of community assembly by successive invasion. Each species is represented by its dynamic population biomass, and a set of fixed traits that determine food-web structure and physiological parameters. Ontogenetic growth is modelled implicitly through wide predator-prey size-ratio windows \citep{rossberg_2012}. The behaviour of consumer individuals is modelled implicitly through Type 2 functional responses extended to incorporate prey switching. Recent variants of the model were described by \citet{fung_pdmm} and \citet{rossberg_2013}.

\subsubsection{Introductory example}
PDMM was able to predict recovery times for Large fish indicator (LFI), Typical length (TyL), Fish population biomass trends (FPBT), Zooplankton to phytoplankton biomass (Z:P) and Zooplankton biomass (ZB). Therefore
\begin{equation*}
M_{PDMM}=\begin{pmatrix}
  0 & 1 & 0 & 0 & 0 & 0\\
  0 & 0 & 1 & 0 & 0 & 0\\
  0 & 0 & 0 & 1 & 0 & 0\\
  0 & 0 & 0 & 0 & 1 & 0\\
  0 & 0 & 0 & 0 & 0 & 1
 \end{pmatrix}.
\end{equation*}

\section{Prior distributions for case study}
\label{app:prior}

\subsubsection*{Long-term individual discrepancy}
\begin{figure}
\begin{center}
\includegraphics[scale=0.7]{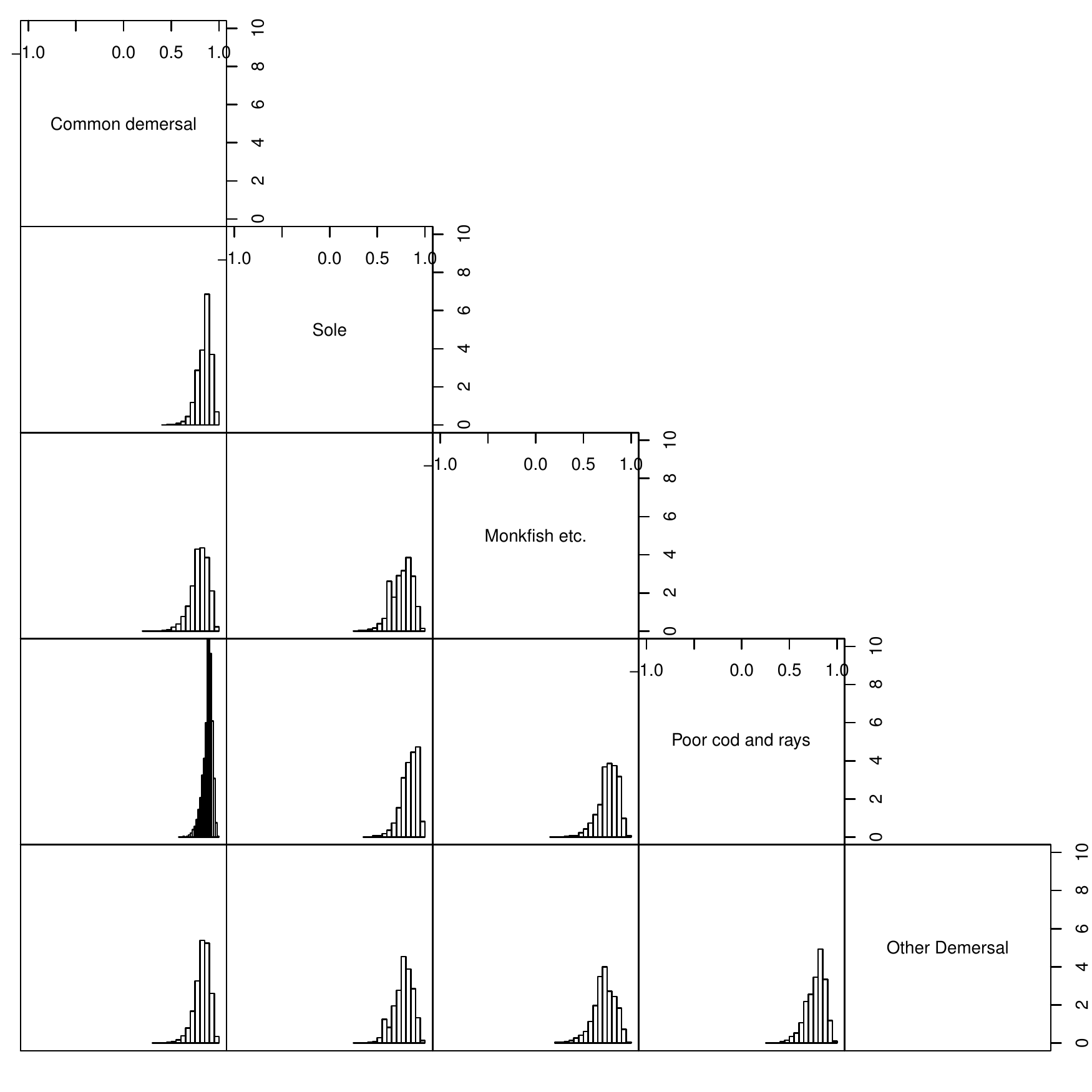}
\end{center}
\caption{Samples from the elicited prior distribution of $P_c$.}
\label{fig_sam:cor}
\end{figure}

JLB decided that she knew more about the correlations of the outputs between simulators than she did the absolute differences. Therefore, using the separation method of \citet{barnard}, the variance of the long-term individual discrepancy, $C$, was rewritten as 
\begin{equation*}
\Sigma_cP_c\Sigma_c
\end{equation*}
where $\Sigma_c$ is a matrix with the diagonal elements $\sigma_j>0$ and the off diagonal elements 0 and $P_c$ is a correlation matrix where the diagonal elements are 1 and the off diagonal elements are $-1<\rho_{cij}<1$. 
The prior on $P_c$ took the form
\begin{equation*}
p(P_c)=\mathbbm{1}_{P_c\succ0}\prod_{i=1}^{n-1}\prod_{j=i+1}^nf_{ij}(\rho_{cij})
\end{equation*}
where $\rho_{cij}$ is the element of $P_c$ on the $i$th row and $j$th column and $f_{ij}$ is a distribution determined by elicitation. Using the method of concordance \citep{gokhale_press,clemen_reilly} and Shelf R package \citep{oakley_shelf}, beta distributions were fitted to JLB's prior beliefs about the elements of $P_c$. These are shown in Table \ref{tab:elic_results} and samples of the off diagonal elements of $P_c$ are shown in Figure \ref{fig_sam:cor}.

\begin{table}[h] \centering 
  \caption{Beta values found from the elicitation} 
  \label{tab:elic_results} 
\begin{tabular}{@{\extracolsep{5pt}} cccccc} 
\\[-1.8ex]\hline 
\hline \\[-1.8ex] 
&\begin{turn}{-300} Common Dem \end{turn}&\begin{turn}{-300}  Sole \end{turn}&\begin{turn}{-300} Monkfish etc. \end{turn}&\begin{turn}{-300}  Poor cod and rays \end{turn}&\begin{turn}{-300}  Other Dem\end{turn}\\
Common Dem&$-$ & $-$ & $-$ & $-$ & $-$ \\ 
Sole&$15.865$,$1.148$ & $-$ & $-$ & $-$ & $-$ \\ 
Monkfish etc.&$15.075$,$1.045$  & $14.168$, $2.085$ & $-$ & $-$ & $-$ \\ 
Poor cod and rays&$53.996$,$4.711$ & $19.115$,$1.628$ & $7.260$,$1.214$ & $-$ & $-$ \\ 
Other Dem&$24.853$,$2.240$ & $14.295$,$1.706$ & $7.050$,$1.549$ & $5.712$,$1.071$ & $-$ \\ 
\hline \\[-1.8ex] 
\end{tabular} 
\end{table} 

\begin{figure}
\begin{center}
\includegraphics[scale=0.7]{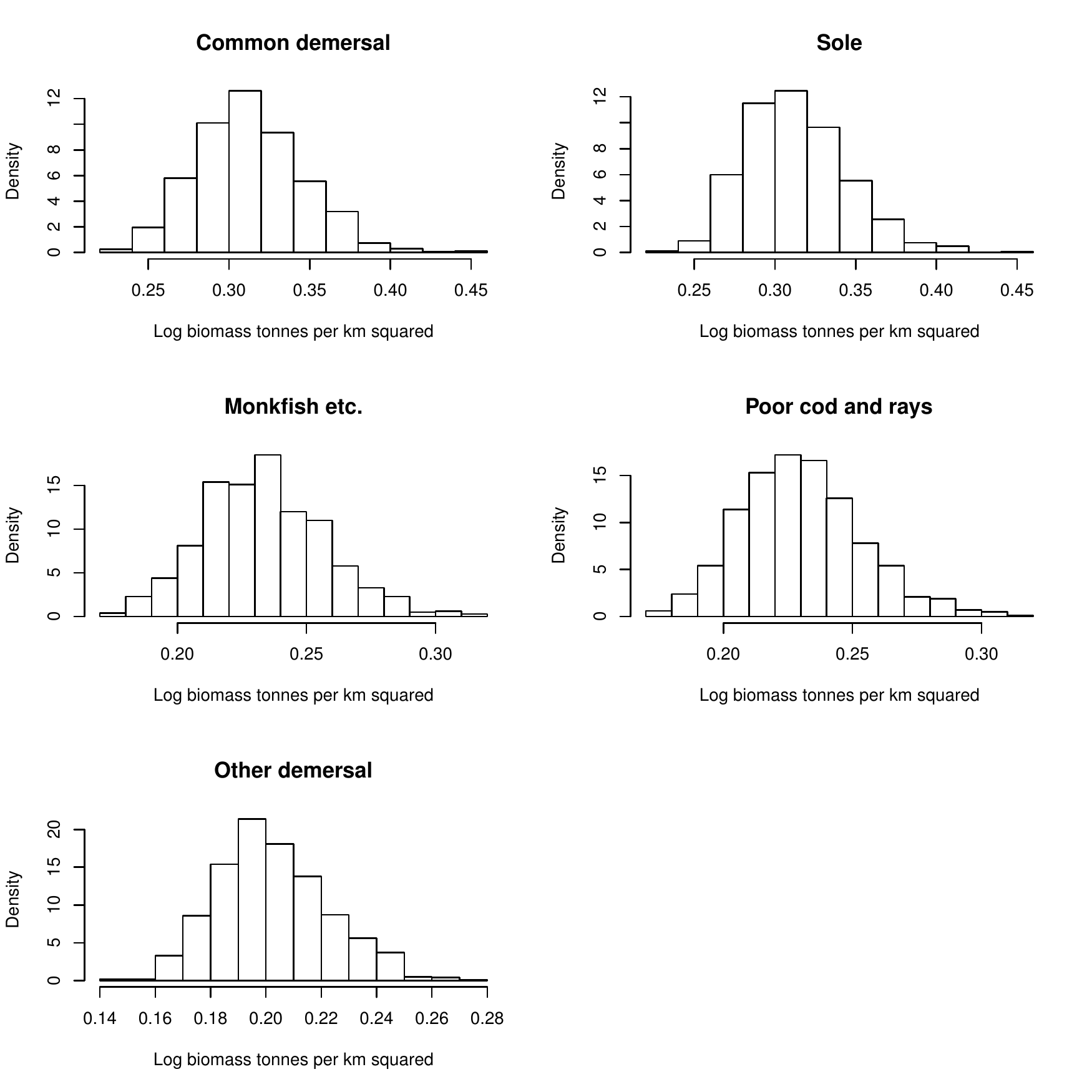}
\end{center}
\caption{Samples from the prior of diagonal elements of $\Sigma_C$.}
\label{fig:var_10}
\end{figure}

Using the difference method of \citet{al_awandhi}, we elicited JLB's beliefs of the absolute values of the simulator outputs for each of the groups species, i.e. the diagonal elements of $\Sigma_C$. All were given  inverse gamma distributions with parameters shown in Table \ref{tab:mu_10}.
\begin{table}[!htbp] \centering 
  \caption{The prior parameters of the inverse-gamma distribution for $\bm\delta$.} 
  \label{tab:mu_10} 
\begin{tabular}{@{\extracolsep{5pt}} ccc} 
\\[-1.8ex]\hline 
\hline \\[-1.8ex] 
&$a$&$b$\\
Common Dem & $100.000$ & $30.801$ \\ 
Sole& $100.000$ & $30.943$ \\ 
Monkfish etc.&$100.000$ & $23.009$ \\ 
Poor cod and rays&$100.000$ & $22.763$ \\ 
Other Dem& $100.000$ & $20.110$ \\ 
\hline \\[-1.8ex] 
\end{tabular} 
\end{table} 
JLB was quite uncertain about the values of $\bm\mu^{(2010)}$ so for numerical stability we capped $a$ to be 100. Samples from the variance, the diagonal elements of $\Sigma_C$, are shown in Figure \ref{fig:var_10}.

\subsubsection*{Shared discrepancy}

JLB did not believe that the model average overestimated the truth any more than it underestimated it so $\bm\delta\sim{}\dist{N}(\bm{0},\mathbb{I})$, where $\mathbb{I}$ is the identity matrix. She was not sure what the standard deviation in equation \ref{eq:zeta_dyn} was so she put quite an uninformative prior,
\begin{equation*}
(diag(\Delta_{\zeta}))^{0.5}\sim{}\dist{Exponential}(5),
\end{equation*}
and the correlation terms were uniformly distributed between -1 and 1. She thought that the standard deviation in equation \ref{eq:mu_dyn} was a bit smaller but was still quite unsure about its value so
\begin{equation*}
(diag(\Delta_{\mu}))^{0.5}\sim{}\dist{Exponential}(1),
\end{equation*}
with correlation terms were uniformly distributed between -1 and 1. $k_{\mu}$, the rate at which the covariance decreases was uniformly distributed on $[0,6]$.

\subsubsection*{Hyper-parameters}

\begin{figure}
\begin{center}
\includegraphics[scale=0.7]{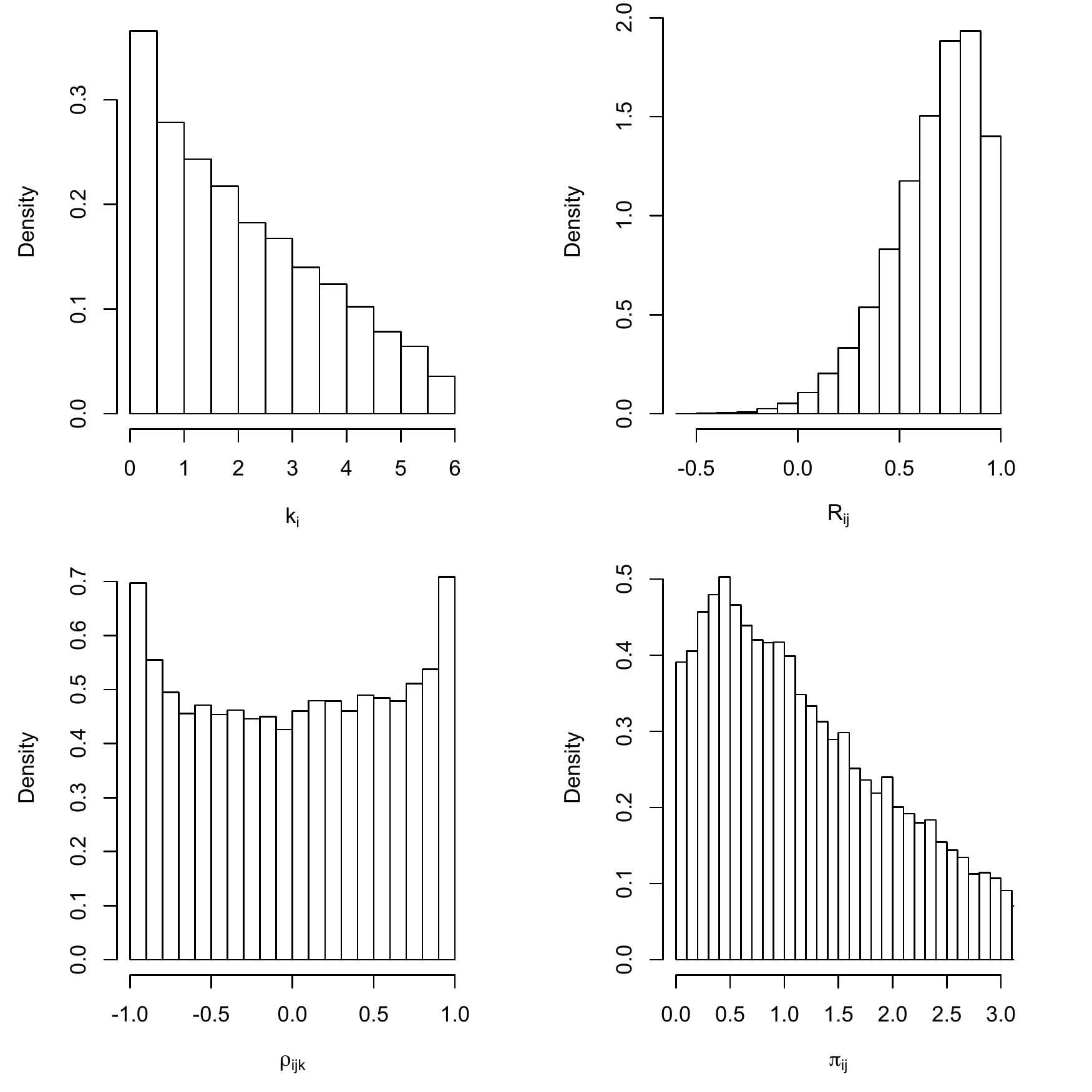}
\end{center}
\caption{Samples from the elicited hyper-priors.}
\label{fig_sam:hyper_prior}
\end{figure}

JLB elicited her prior beliefs for the hyper-parameters $k_i$, $R_{ij}$, $\pi_{ij}$ and $\rho_{ijk}$ for $j=1,\ldots,5$, $k=1,\ldots,5$ and $j\neq{}k$. She believed that for all $j$ and $k$ the distributions were exchangeable a priori, thus the prior predictive distributions were the same for all $j$ and $k$. Samples from the prior predictive distributions are shown in Figure \ref{fig_sam:hyper_prior}.
The hyper-priors were tuned so that they would lead to prior predictive distribution of JLB shown in Figure \ref{fig_sam:hyper_prior}.

JLB expected that models would reach equilibrium slower than in one step. However she was not quite sure how fast that would happen so she set quite an uninformative priors on $k_i$. By thinking about the rate by which the variance declines, $\exp{{k_i}}$, the priors were elicited to be
\begin{equation*}
a_k\sim{}\dist{Gamma}(10,10)
\end{equation*}
and
\begin{equation*}
b_k\sim{}\dist{Gamma}(10,5).
\end{equation*}
A priori, JLB did not know what values the diagonals of $R_i$ but suggested that they would be positive so
\begin{equation*}
a_R\sim{}\dist{Gamma}(10,5)
\end{equation*}
and
\begin{equation*}
b_R\sim{}\dist{Gamma}(10,10).
\end{equation*}
Similarly, JLB expected $\pi_j$ to be small. Therefore
\begin{equation*}
\alpha_{\pi,j}\sim{}\dist{Gamma}(15,10)
\end{equation*}
and
\begin{equation*}
\alpha_{\pi,j}\sim{}\dist{Gamma}(10,2).
\end{equation*}
JLB was unsure about values of $\rho_{jk}$ and therefore
\begin{equation*}
a_{\rho{}jk},b_{\rho{}jk}\sim{}\dist{Gamma}(10,10).
\end{equation*}

\section{Correlation of the future groups}
\label{app:res}

The correlations of the relative truth of the future groups in 2025 and 2050 are shown in Tables \ref{tb:cor2025} and \ref{tb:cor2050} respectively.

\begin{table}[h] \centering 
  \caption{The correlation of the relative truth for the future groups in 2025.} 
  \label{tb:cor2025} 
\begin{tabular}{@{\extracolsep{5pt}} cccccc} 
\\[-1.8ex]\hline 
\hline \\[-1.8ex] 
&\begin{turn}{-300} Common Dem \end{turn}&\begin{turn}{-300}  Sole \end{turn}&\begin{turn}{-300} Monkfish etc. \end{turn}&\begin{turn}{-300}  Poor cod and rays \end{turn}&\begin{turn}{-300}  Other Dem\end{turn}\\
Common Dem&$1$ & $0.156$ & $0.281$ & $0.263$ & $0.023$ \\ 
Sole&$0.156$ & $1$ & $0.285$ & $0.134$ & $0.001$ \\ 
Monkfish etc.&$0.281$ & $0.285$ & $1$ & $0.290$ & $$-$0.017$ \\ 
Poor cod and rays&$0.263$ & $0.134$ & $0.290$ & $1$ & $$-$0.025$ \\ 
Other Dem&$0.023$ & $0.001$ & $$-$0.017$ & $$-$0.025$ & $1$ \\ 
\hline \\[-1.8ex] 
\end{tabular} 
\end{table} 

\begin{table}[h] \centering 
  \caption{The correlation of the relative truth for the future groups in 2050.} 
  \label{tb:cor2050} 
\begin{tabular}{@{\extracolsep{5pt}} cccccc} 
\\[-1.8ex]\hline 
\hline \\[-1.8ex] 
&\begin{turn}{-300} Common Dem \end{turn}&\begin{turn}{-300}  Sole \end{turn}&\begin{turn}{-300} Monkfish etc. \end{turn}&\begin{turn}{-300}  Poor cod and rays \end{turn}&\begin{turn}{-300}  Other Dem\end{turn}\\
Common Dem&$1$ & $0.138$ & $0.306$ & $0.172$ & $0.011$ \\ 
Sole&$0.138$ & $1$ & $0.273$ & $0.154$ & $0.011$ \\ 
Monkfish etc.&$0.306$ & $0.273$ & $1$ & $0.253$ & $0.023$ \\ 
Poor cod and rays&$0.172$ & $0.154$ & $0.253$ & $1$ & $$-$0.010$ \\ 
Other Dem&$0.011$ & $0.011$ & $0.023$ & $$-$0.010$ & $1$ \\ 
\hline \\[-1.8ex] 
\end{tabular} 
\end{table}

\end{document}